\documentclass[fleqn,usenatbib]{mnras}

\usepackage[T1]{fontenc}
\usepackage{ae,aecompl}

\usepackage{graphicx}	
\usepackage{amsmath}	
\usepackage{amssymb}	

\usepackage[normalem]{ulem}
\usepackage{ragged2e}
\usepackage{xcolor}
\usepackage{color}

\usepackage{newtxtext,newtxmath}

\title[IMBH binaries in nucleated dwarf galaxies]{Extremely efficient mergers of intermediate mass black hole binaries in nucleated dwarf galaxies}

\author[Khan and Holley-Bockelmann]{Fazeel Mahmood Khan$^{1,2,3}$\thanks{E-mail: khanfazeel.ist@gmail.com}
	and
	Kelly Holley-Bockelmann$^{1,4}$
	\\
	$^{1}$Department of Physics and Astronomy, Vanderbilt University, Nashville, TN 37240, USA\\
	$^{2}$Department of Space Science, Institute of Space Technology, Islamabad 44000, Pakistan\\
	$^{3}$Space and Astrophysics Research Lab (SARL), National Centre of GIS and Space Applications (NCGSA), Islamabad 44000, Pakistan\\ 
	$^{4}$Department of Physics, Fisk University, Nashville, TN 37208, USA\\
}
\date{Accepted XXX. Received YYY; in original form ZZZ}

\pubyear{2016}

\begin{document}
	\label{firstpage}
	\pagerange{\pageref{firstpage}--\pageref{lastpage}}
	\maketitle
	
	\begin{abstract}
		
		Gravitational waves emitted by merging black holes between $\sim 10^4-10^7~M_\odot$ will be detectable by the Laser Interferometer Space Antenna (LISA) with signal-to-noise ratios of several hundred out to redshift 20. Supermassive black hole ($10^7$~M$_{\odot}$ - $10^{10}$~M$_{\odot}$) binary formation, coalescence and merger within massive galaxies is well-studied. However, these low-to-intermediate mass black holes (IMBHs) are hosted by low-mass and dwarf galaxies; it is not trivial to extrapolate black hole merger timescales to this IMBH binary regime, due to the starkly different host galaxy structure, kinematics, and morphology compared to massive galaxy hosts. We perform ultra-high resolution $N$-body simulations to study IMBH dynamics in nucleated dwarf galaxies whose structural parameters are obtained from observations of nearby dwarf galaxies. Starting from 50 parsecs, an IMBH quickly forms a binary. Thereafter, the binary orbit shrinks rapidly due to the high central stellar densities furnished by nuclear star clusters (NSCs). We find high eccentricities ($e \sim 0.4-0.99$) in our suite of IMBH binaries, and residual eccentricity may persist to the LISA regime. IMBH merger times are typically a few hundred million years, with a few exceptionally short merger times for high eccentricities. We find that IMBH-stellar encounters originate predominantly from NSCs, if the NSC-to-IMBH binary mass ratio is greater than 10; otherwise, bulge stars contribute significantly.  As the IMBH binary ejects stars, however, the NSCs is disrupted. We conclude that comparable-mass IMBHs merge very efficiently in nucleated dwarf galaxies, making them promising LISA sources, as well as a channel for IMBH growth.

	\end{abstract}


	

	\section{Introduction}\label{sec-intro}
	
	The overwhelming majority of galaxies with stellar mass $10^8 - 10^{11}$ $M_{\odot}$ host a nuclear star cluster (NSC), a massive, dynamically distinct and incredibly dense system of stars at the heart of a galaxy~\citep{neu20}. NSCs correlate with their host galaxy properties, such as mass, luminosity, and effective radius~\citep{sco13,den14,san19a}. Kinematic studies suggest that NSC rotation is common, with $v/\sigma$ approaching $\sim 1$ in many cases\citep{seth08,seth10,feld14,ngu18}. With the deep potential well and unique dynamical environment in an NSC, conditions are ripe for stellar collisions, tidal disruption events (TDEs), few-body interactions, and perhaps the stellar origin binary black hole mergers seen by LIGO/Virgo.

	It was once thought that NSCs took the place of massive central black holes, but it is becoming more clear that this not the case. The problem is that in these comparatively low-mass galaxies, black hole mass measurements are extremely challenging due to the weak gravitational impact they cause on their surroundings.
	However, advancements in detailed modeling, integral field spectroscopy and high spatial resolution observations have led to breakthroughs in uncovering tentative evidence of a population of $10^4$--$10^{6}$~M$_{\odot}$ black holes; we will refer to these as intermediate mass black holes (IMBH). Indeed, a growing number of IMBH candidates are reported in low-mass and dwarf galaxies \citep{rei13,ngu17,mez18,ngu18}, and as these detections increase, a picture is emerging that IMBHs and NSCs co-exist and may co-evolve. 
	
	Dwarf and low-mass galaxy evolution is expected to be punctuated by galaxy mergers and interactions. Although the galaxy-galaxy merger rate for low mass galaxies is not as well-measured, the consensus is that the major merger rate for galaxies with stellar mass $\sim 10^8 \, M_\odot$ (0.02 per galaxy per Gyr) is about 1/3 that of the most massive galaxies in the local universe~\citep[e.g.][]{Casteels2014, Duncan2019, Ventou2019}. Such galaxy mergers should give rise to binary IMBHs, which, much like their more massive supermassive black hole (SMBH) counterparts, should merge and produce loud gravitational wave signals in the LISA band \citep{amaro+17}.	

	On the high mass end of the black hole mass spectrum, the formation and dynamical evolution of SMBH binaries within a galactic center has been studied extensively over the last 15 years, both with purely stellar \citep{berczik+06,just11,khan+11,preto+11,Gualandris+12,khan+12a,vasiliev+15,bor16,kha16} and purely gas dynamics \citep{esc05,dotti07,nixon11,mayer13,roe14}.
	These studies conclude that there is rich and diverse evolution of SMBH binaries that depends on the morphology, structure, and kinematics of their hosts, as well as on the initial orbit of the infalling SMBH.
	
	At the upper end of black hole mass function, SMBHs are hosted by giant elliptical galaxies having shallow stellar density profiles, and these tend to take longer time to merge $\sim$ few Gyr. On the other hand, SMBHs in Milky Way mass galaxies tend to merge faster, $\sim$ few hundred Myrs, thanks to higher stellar density in these nuclei. The eccentricity behaviour in these two classes of hosts is also quite different, with binaries in giant galaxy mergers possessing very high eccentricities approaching unity, and binaries in lower mass galaxy mergers evolving with intermediate to low eccentricities \citep{khan+12b,ran17,khan18a}.

	In SMBH host galaxies, rotation of some degree is ubiquitous~\citep{Cappellari+16}, and it has been shown that rotating stellar environments add complexity to the binary evolution.  Binaries in co-rotation with their galaxy hosts tend to shrink faster due to a longer interaction time between a typical star and the binary. In this case, the binary eccentricity is largely circular, but astoundingly, the center of mass of the {\it binary itself} settles into a prograde orbit, roughly the size of the binary's influence radius \citep{Mirza+17}. SMBH binaries that are counterrotating with respect to the stellar surrounding will initially shrink at a slower pace from dynamical friction, but acquire very high eccentricities in the hard binary phase that speeds up the final coalescence \citep{Sesana+11,holley+15,Mirza+17}. Along the way, a SMBH binary may flip its orbital plane to align with the rotating stellar background. 
	
	Although the dynamics of SMBHs are relatively well studied, there is much to explore at the lower mass end. 
	IMBH pairs in a dwarf galaxy nucleus can assemble either by mergers of dwarf galaxies or through migration of globular clusters, if at least some of them happen to host IMBHs \citep{KHB08}. We note that the globular cluster accretion scenario has long been proposed to form NSCs \citep{tre75,har11,ant12}. 
	Broadly, the evolution of IMBH binaries in dwarf galactic nuclei is expected to happen in similar phases as for SMBH binaries, but the timescales and binary parameter evolution could be very different due to important differences in the galaxy host structure and kinematics, namely the incredibly dense NSC environment. If IMBHs have a similar occupation fraction in low mass galaxies as SMBHs do in high mass hosts, it has important implications for the LISA merger rate, as low mass galaxies dominate the number density.  
	
	Intuiting the IMBH merger rate density is not clear cut, however; although we have measured the {\it galaxy} merger rate per volume \citep{Casteels2014}, the lower escape velocity of these hosts may make IMBHs particularly vulnerable to gravitational wave recoil \citep{KHB08,gua08}, yet their low mass may make it more difficult to pair the IMBHs in the first place. It is thus of paramount importance to understand IMBH orbital dynamics and merger timescales to produce reliable forecasts of merger rates.
	
	This study investigates IMBH pair evolution and merging in equilibrium models of dwarf galaxy nuclei by performing a suite of extremely high resolution direct $N-$body simulations. In an effort to reproduce faithful dwarf galaxy realizations, we generate galaxy models having the same structural, morphological, kinematic and IMBH characteristics as observed in nearby dwarf galaxies \citep{ngu18}. These include accurate masses of the bulge central black hole and bulge, the Sersic indices of the bulge and NSCs, their effective radii and flattening ratios. As we mentioned above that the rotation of the stellar component has profound effects on MBH binary evolution, we also use observations of dwarf galaxy kinematics down to the inner parsec to initialize our galaxy models; this will explore the effect of rotation on IMBH mergers in this lower mass regime. This is the first direct N-body study of its kind for two reasons: 1) we explore an entirely different regime of IMBHs and their low mass hosts; and 2) our simulations have an unprecedented degree of realism because the models are constructed from specific dwarf galaxy observations. 
	
	There is an increasing body of work on the dynamics of low mass black holes, in both isolated galaxy interactions and in zoom-in cosmological simulations. For example, \citet{tam18} performed N-body simulations of isolated dwarf galaxy mergers mainly to quantify the effect of the dark matter halo on the formation of a hard binary IMBH. They found that cuspy dark matter halos were critical in ushering the infalling IMBH to the center, but were unable to follow the binary evolution further, due to the inherent softening and force resolution limitations of tree-like codes. In addition, this study did not model nucleated dwarfs, which was noted in the paper as a potential caveat.  
	Exploring the co-evolution of low mass galaxy and their hosts in a cosmological context, ~\citet{Bellovary2019} discovered a large population of IMBHs wandering in the outskirts of a galaxy as a result of multiple galaxy mergers and long IMBH orbital decay times. Black hole mergers, when they occurred, had very unequal masses, with mass ratios of 1:50 being typical. These qualitative conclusions were borne out in earlier work~\citep{Micic,Bellovary2019,KHB2010}, though we caution that the force resolution of even tour-de-force cosmological simulations may be too coarse to capture the true orbital evolution of IMBHs. Indeed, even with a careful treatment of dynamical friction designed to mitigate the inaccurate orbital evolution caused by a coarse-grained potential, these treatments rely on interpolating the unresolved density from the particle distribution, and since most cosmological simulations are not able to accurately resolve structures as compact and dense as NSCs, the black hole dynamics may still be inaccurate. \citet{ogi19} performed isolated merger simulations of nuclear star clusters each harboring a central MBH and concluded that MBH merger times may vary from a few hundred million years to 5 billion years, strongly depending on the mass ratios of merging MBHs. However, these findings are less reliable, both due to the extremely small particle number adopted in their simulation suite and, as we will show, the absence of the stellar bulge surrounding the NSCs.

	This paper is organised as follows: Section 2 describes the observed key parameters of the dwarf galaxies comprising our simulation suite, together with our model generation scheme and stability analysis. Section 3 presents the results of our simulations in detail, focusing on IMBH binary evolution and its impact on the surrounding stellar shroud. Section 4 summarizes and discusses implications for future gravitational wave observations.

	\section{Galaxy Sample, Initial Setup and Numerical Methods} \label{sec:galaxies} 
	\begin{figure*}
		\centerline{
			\resizebox{0.95\hsize}{!}{\includegraphics[angle=270]{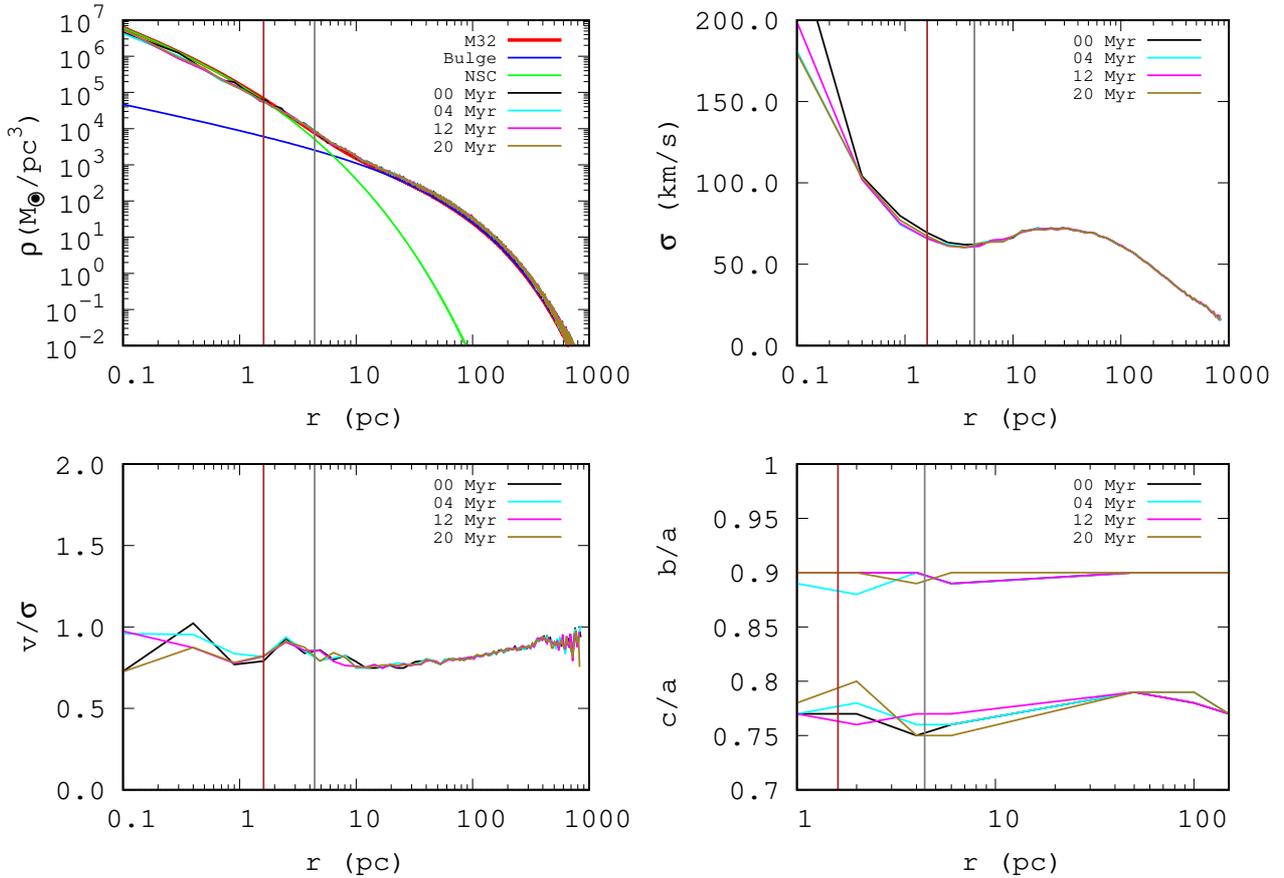}}
		}
		\caption{
			Our M32 analog based on observational parameters (red line) and after (lines labeled by Myr) stability analysis. Vertical lines in each panel show the IMBH influence radius (brown, left) and NSC effective radius (grey, right).  Top left: Stellar density profiles. Top right: velocity dispersion $\sigma$ at various times. Bottom left: The ratio of $v/\sigma$. Bottom right: Intermediate to major (b/a) and minor to major (c/a) axes ratio profiles. 
		} \label{fig:m32multi}
	\end{figure*}
	On the high mass end of the black mass spectrum, it is now understood that SMBH binary evolution depends critically on the structure of the host galactic nucleus. Observations of galaxy parameters such as the black hole influence radius, the shape of density profile, effective radius, rotation curves and dispersion profiles, if available, help build appropriate models to study SMBH pair evolution in realistic stellar surroundings. 
	In this study, we investigate IMBH binary evolution in 5 nearby dwarf galaxies whose IMBHs and structural properties are obtained from high resolution stellar photometric and kinematic observations of \citet{ngu17,ngu18}.

	\citet{ngu18} describe the surface brightness of dwarfs with Sersic profiles \citep{ser63,ser68}, and
	we transform the Sersic profile to a 3D stellar density profile using \citet{pru97} (see also \citet{ter05}): 
	(equation \ref{eq:sersic}). 
	
	\begin{equation}
	\rho(r) = \rho_0 \left( \frac{r}{R_e} \right)^{-p} e^{-b(r/R_e)^{(1/n)}},
	\end{equation} \label{eq:sersic}
	
	\noindent where $r$ is the radial distance from the galaxy center, n is Sersic index ,and $R_e$ is effective radius. $\rho_0$ is normalization density that depends on total galaxy mass and effective radius. The power-law index, $p$, and the $b$ parameter can be expressed as: $p = 1.0-0.6097/n + 0.055 63/n^2$,  and $b \approx 2\,n-1/3+0.009876/n$, for $\sim 0.5 < n < 10$  \citep{ter05}.
	
	
	\begin{figure*}
		\centerline{
			\resizebox{0.95\hsize}{!}{\includegraphics[angle=270]{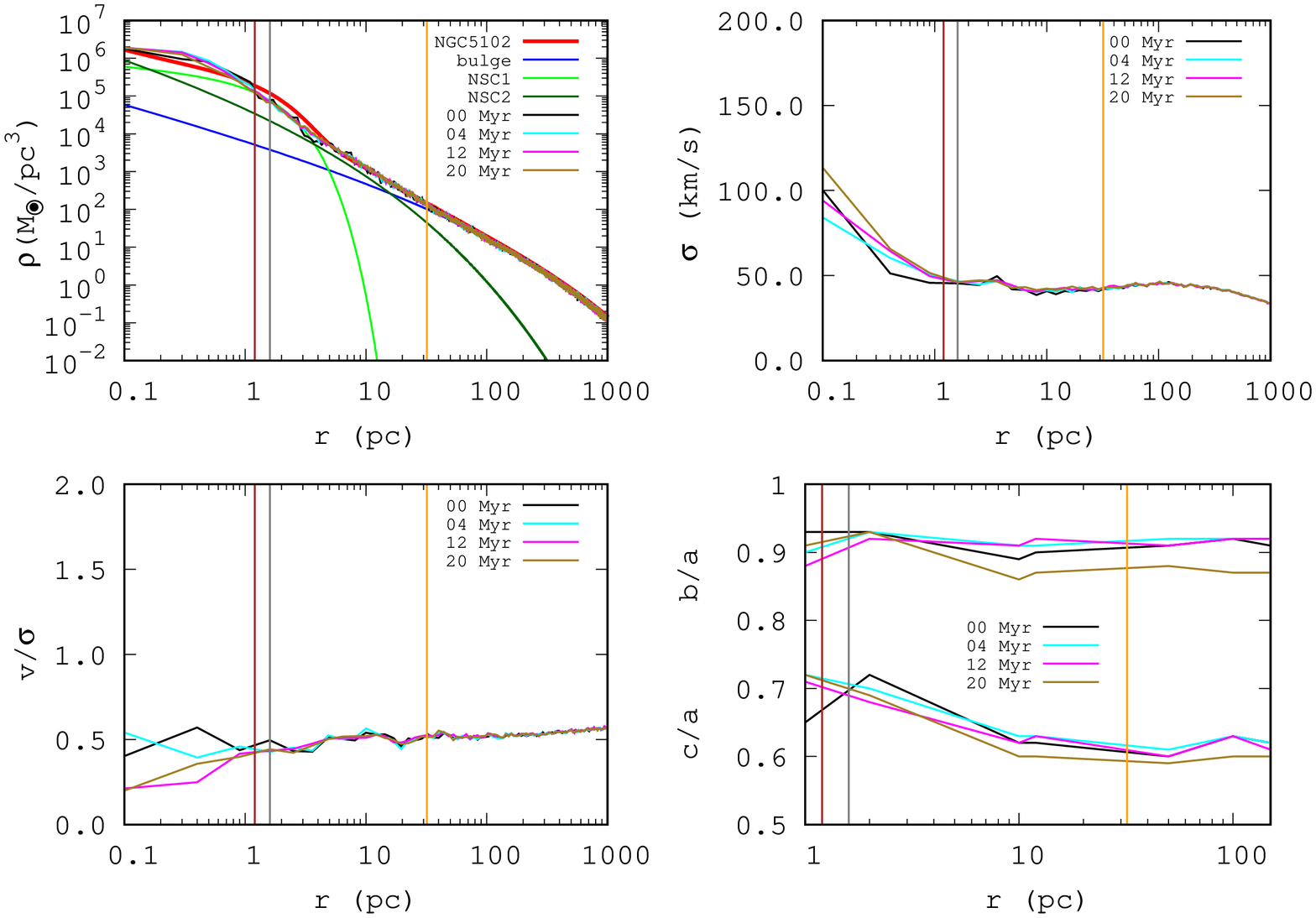}}
		}
		\caption{
			Our initial model and stability analysis for NGC 5102. The panels, labels and lines are the same as in figure \ref{fig:m32multi}. However, there is an additional vertical orange line marking the effective radius of the second NSC. 
		} \label{fig:ngc5102multi}
	\end{figure*}
	Since there is an ambiguity in translating the 2-d axial ratio measurements (denoted here as $a_2/a_1$) into a 3-d shape (denoted as $b/a; c/a$), we choose 3-d shapes that are very conservatively triaxial and close to oblate. This is broadly consistent with observations of 3-d shape measurements of elliptical and dwarf galaxies \citep{Kormendy+13,san19a}, and is similar to the shapes adopted in prior studies of MBH merging timescales, which will better allow us to compare these results to what is found in more massive ellipticals. Key observational parameters of the sample are presented in Tables 2-6.
	
	Galaxies are then realised using open source modelling software \textit{smile} \citep{vasi13}. \textit{smile} can build multi-component galaxy models in dynamical equilibrium with desired shape parameters using Schwarzschild's orbit superposition method \citep{sch79}. There are several options to construct an N-body model within \textit{smile}; we adopted the Eddington method with 20 radial and 6 angular segments, but explored minor variations to this procedure and found no significant difference in model stability. We intend to explore the effect of chaotic orbit fraction in the black hole merging timescale, but since direct N-body simulations are computationally expensive, we reserve a comprehensive study for a future paper. Rotation, if required to match the observed rotation profile, is introduced by flipping the z-component of angular momentum of a fraction of particles initially. For this study, we generate galaxy models with 2 million particles, except for NGC 205, which required 4 million particles to better resolve the low IMBH mass. This results in a varying mass and spatial resolution for each model, noted in table \ref{tab:simparam}.
	
	Orbit modeling is a fast and elegant method to populate a galaxy, but it is important to keep in mind that the set of orbits selected in a potential-density pair is not a unique solution to the galaxy model. Moreover, there is not guarantee that a resulting model will be stable -- this is particularly true of black hole-embedded potentials~\citep{Valluri98, KHB02}.  To ensure that the initial conditions we generate do preserve their shape, structure, and kinematics before a second IMBH is introduced, we run $N-$body simulations of each model for 20 Myr using $\phi-$GPU (section \ref{subsec:phigpu}). For almost all the runs in our suite, an IMBH binary forms before 20 Myr and further evolution of the stellar profile beyond this time would be caused by massive binary. 
	
	\subsection{$\phi$-GPU} \label{subsec:phigpu}
	
	This suite of simulations is performed using  $\phi$-GPU \citep{berczik+11}, a massively parallel direct $N-$body code employing the power of Graphical Processing Units (GPUs) to calculate pairwise forces between all the particles. $\phi$-GPU employs softening in the force calculations, and the current version of code handles individual softening for each particle \citep{li12,Wang+14,li17}. Stellar particles are assigned a softening of $ \epsilon_{\star} = 10^{-4}$ and black holes are given $\epsilon_{IMBH}=0$ softening in model units. We investigated the effect of $\epsilon_{\star}= [10^{-6}$ -- $10^{-3}]$ and found that the results are convergent for a value of $\leq 10^{-4}$.  Softening in physical units depends on our choice of the length unit [LU] in each galaxy model, but 1 LU is set to be roughly effective radius of the bulge, which is typically of the order of 100 pc in this sample. This results in a typical star-star interaction softening of 0.01 pc, which is sufficient to discourage stellar binary formation.  Star-IMBH softening is calculated using 
	
	\begin{equation}
	\epsilon_{\star,IMBH} = 0.1 \times \sqrt{\frac{\epsilon_{\star}^2 + \epsilon_{IMBH}^2}{2}},
	\end{equation} \label{eq:soft}
	
	\noindent resulting in $\epsilon_{\star,IMBH} = 7 \times 10^{-6} $ in model units. In physical units, it corresponds to a typical value of $\sim 7 \times 10^{-4}$ pc. Values for these parameters in each individual model are described in table \ref{tab:simparam}.
	
	\begin{table*}
		\begin{center}
			\vspace{-0.5pt}
			\caption{Simulation Parameters} 
			\begin{tabular}{c c c c c c c}
				\hline
				Galaxy & $N$ & MU ($10^9$ M$_{\odot}$) & LU (pc) & M$_{IMBH}$/$M_{\star}$ & $\epsilon_{\star,\star}$ & $\epsilon_{\star,IMBH}$\\
				\hline
				M32 & $2$ & $0.99$ & $100$ & $714$ & $0.01$ & $0.0007$\\
				NGC5102 & $2$ & $6.0$ & $100$ & $293$ & $0.01$ & $0.0007$\\
				NGC5206 & $2$ & $2.4$ & $100$ & $392$ & $0.01$ & $0.0007$\\
				NGC404 & $2$ & $0.86$ & $100$ & $163$ & $0.01$ & $0.0007$\\
				NGC205 & $4$ & $0.97$ & $100$ & $160$ & $0.01$ & $0.00007$\\
				\hline
			\end{tabular}
			
			\vspace{15pt}
			Column~1: Galaxy name. Column~2: Number of particles. Column~3: Mass unit in our model. Column~4: Length unit in our model. Column~5: Secondary black hole to stellar particle mass ratio. Column~6: Star-star interaction softening. Column~7: Star-IMBH interaction softening. \label{tab:simparam}
			\vspace{15pt}
		\end{center}
	\end{table*}
	
	\subsection{Galaxy sample} \label{galaxy sample}
	
	\begin{figure*}
		\centerline{
			\resizebox{0.95\hsize}{!}{\includegraphics[angle=270]{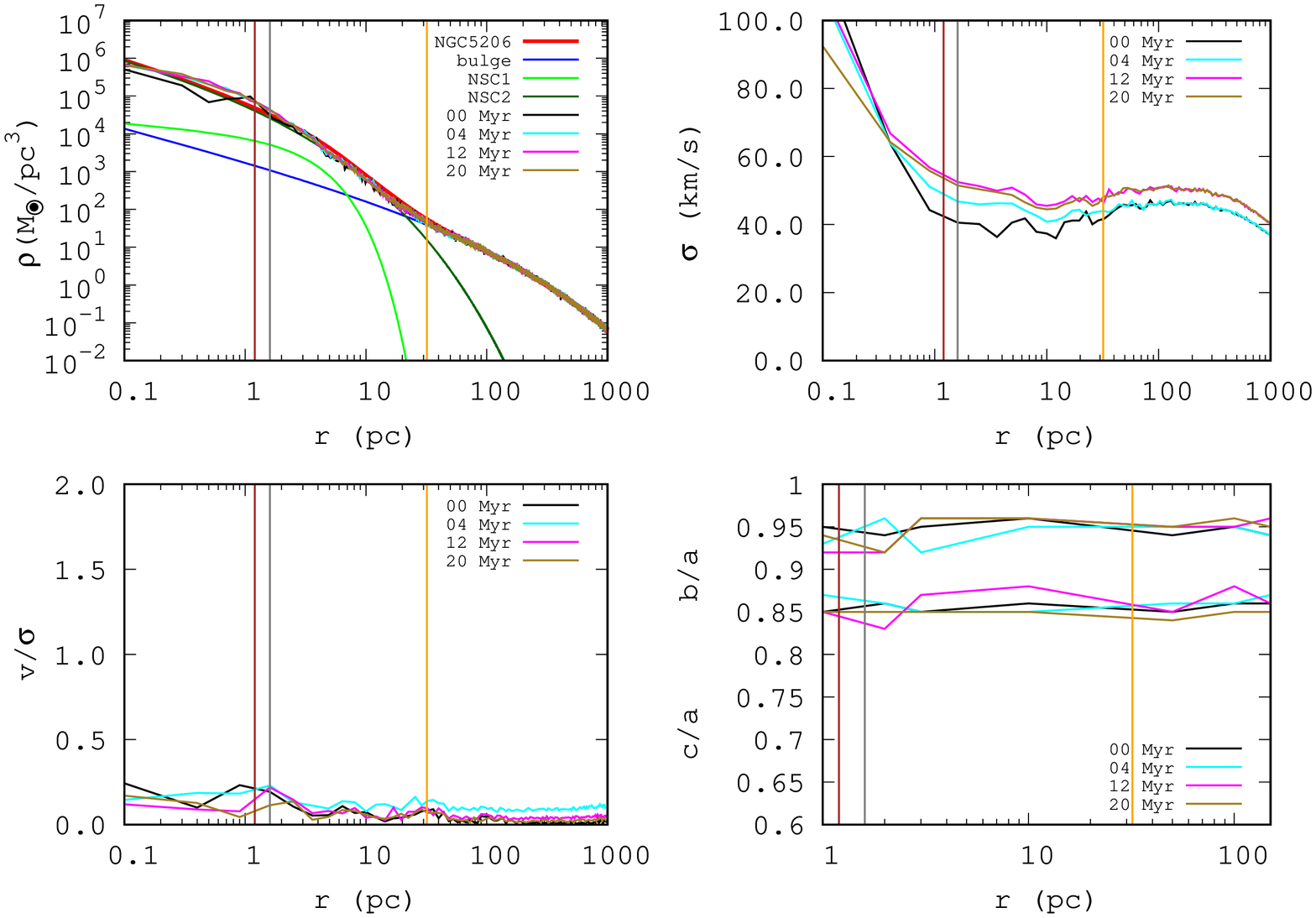}}
		}
		\caption{Our initial model and stability analysis for NGC 5206. The panels, labels and lines are the same as in figure \ref{fig:ngc5102multi}. 
		} \label{fig:ngc5206multi}
	\end{figure*}
	
	Here, we briefly discuss properties of each galaxy in our sample and also describe our initial models and their stability. 
	
	\subsubsection{M32} \label{subsec:m32}
	
	M32 is a dense dwarf elliptical galaxy orbiting the Andromeda galaxy (M31). Its central stellar density reaches $\rho_{\star,c} \simeq 10^{7}$ M$_{\odot}/{\rm pc}^3$ inside central $0.1$~pc \citep{lau98,ver02}, and possesses a prominent nuclear star cluster which hosts a central black hole that is evident from rising velocity dispersion profile. The black hole mass is estimated to be $2.5 \times 10^6$ M$_{\odot}$ with an influence radius $\simeq 1.61$ pc \citep{ngu18}. 
	
	\begin{table}
		\vspace{-0.5pt}
		\caption{M32 Galaxy Parameters} 
		\begin{tabular}{c c c c c}
			\hline
			Component & $n$ & $r_{\rm eff}$/$r_{\rm infl.}$(pc) & $M_{\star}$ ($10^7$ M$_{\odot}$) & $a_2/a_1$\\
			\hline
			IMBH & $--$ & $1.61$ & $0.25$ & $--$\\
			NSC & $2.7$ & $4.4$ & $1.45$ & $0.75$\\
			Bulge & $1.6$ & $108$ & $79.4$ & $0.79$\\
			Disk & $1.0$ & $516$ & $19.3$ & $0.79$\\
			\hline
		\end{tabular}\label{tab:m32param}
		\vspace{15pt}
		
		Column~1: Galaxy component. Column~2: Sersic index. Column~3: Effective radius of the galaxy component (note: for the IMBH, we quote the influence radius instead). Column~4: Stellar mass.  Column~5: 2-d axis ratio as measured by GALFIT \citep{Peng2010}.
		
		\vspace{15pt}
	\end{table}

	Our model is constructed using the bulge and NSC as described in table \ref{tab:m32param} together with a central black hole. We neglect the disk component because it has a larger effective radius and smaller stellar mass compared to bulge and hence has a less significant effect on the black hole pair dynamics, which is mainly governed by stellar parameters near a few influence radii of the black hole ($1.61$ pc in our case). Note that we also neglect the dark matter halo, which would dominate the large scale infall time, but should also be less important in the inner parsec.
	Figure \ref{fig:m32multi} shows the density and kinematic profiles of our M32 galaxy model, and their initial stability.  The brown and gray vertical lines denote the black hole's influence radius and the effective radius of NSC, respectively. We find that the density profile (upper left panel) is very stable and note the clear density enhancement within the inner 4 parsecs due to the presence of the NSC, reaching up to $\simeq 10^{7}$ M$_{\odot}/{\rm pc^3}$. 
	Density profile and velocity profiles globally, at the IMBH radius of influence and at the NSC effective radius are very stable. Inside roughly 0.2 parsecs, the velocity and velocity dispersion both rise, but the change is well within the limits of the observations \citep{ngu18}. $v/\sigma \sim 0.7$ over most of the model is also consistent with observed values. The M32 initial model preserves the intended shape parameters ($b/a \sim 0.9 \mathrm ~{and}~ c/a \sim 0.75-0.8$) over the course of the stability run.

	\subsubsection{NGC 5102} \label{subsec:ngc5102}
	
	NGC 5102 exhibits the morphological attributes of an S0 type galaxy \citep{dav08}. Kinematic measurements yield a rotational velocity of $20$ km/s and a velocity dispersion in range of $44-60$ km/s, with higher values observed near the center~\citep{mit17}. $v/\sigma$ is around $0.4$ close to the center and increases to $0.7$ at larger radii \citep{ngu18}. The kinematic profile suggests that the inner NSC is weakly rotating while the outer NSC is a flattened ($a_2/a_1 \sim 0.6$) and strongly rotating system, so we choose to model these components as two separate NSCs. The estimated mass of the central black hole is $8.8 \times 10^5$ M$_{\odot}$, having an influence radius of $1.2$ pc, and the total stellar mass of the galaxy is estimated to be $6 \times 10^9$ M$_{\odot}$ \citep{ngu18}. Table \ref{tab:ngc5102param} showcases key parameters of the various stellar components of NGC 5102. 
	
	\begin{table}
		\vspace{-0.5pt}
		\caption{NGC 5102 Galaxy Parameters} 
		\begin{tabular}{c c c c c}
			\hline
			Component & $n$ & $r_{\rm eff}$(pc) & $M_{\star}$ ($10^7$ M$_{\odot}$) & $a_2/a_1$\\
			\hline
			IMBH & $--$ & $1.2$ & $0.088$ & $--$\\
			NSC1 & $0.8$ & $1.6$ & $0.71$ & $0.68$\\
			NSC2 & $3.1$ & $32$ & $5.8$ & $0.59$\\
			Bulge & $3.0$ & $1200$ & $592$ & $0.60$\\
			\hline
		\end{tabular}\label{tab:ngc5102param}
		\vspace{15pt}
		
		Column~1: Galaxy component. Column~2: Sersic index. Column~3: Effective radius of the galaxy component (note: for the IMBH, we quote the influence radius instead). Column~4: Stellar mass. Column~5: 2-d axis ratio as measured by GALFIT.
		\vspace{15pt}
	\end{table}

	Figure \ref{fig:ngc5102multi} shows our realisation of the NGC 5102 galaxy and its stability analysis. Parameters presented in figure are consistent with those reported in above mentioned literature as well as reasonably stable at most of spatial scale over the duration of stability run. Well inside the influence radius, the velocity profiles of our galaxy model evolve and $v/\sigma$ drops to about $0.25$. This evolution is consistent with the resolution limits of the observations, which are about a parsec for NGC 5102. Since the majority of the binary black hole loss cone originates from a few times the IMBH influence radius, this inner evolution will not affect the orbital decay time appreciably~\citep{seskha+15}, though we caution that it may have an effect on the eccentricity or center of mass motion of the binary IMBH.

	\begin{figure*}
		\centerline{
			\resizebox{0.95\hsize}{!}{\includegraphics[angle=270]{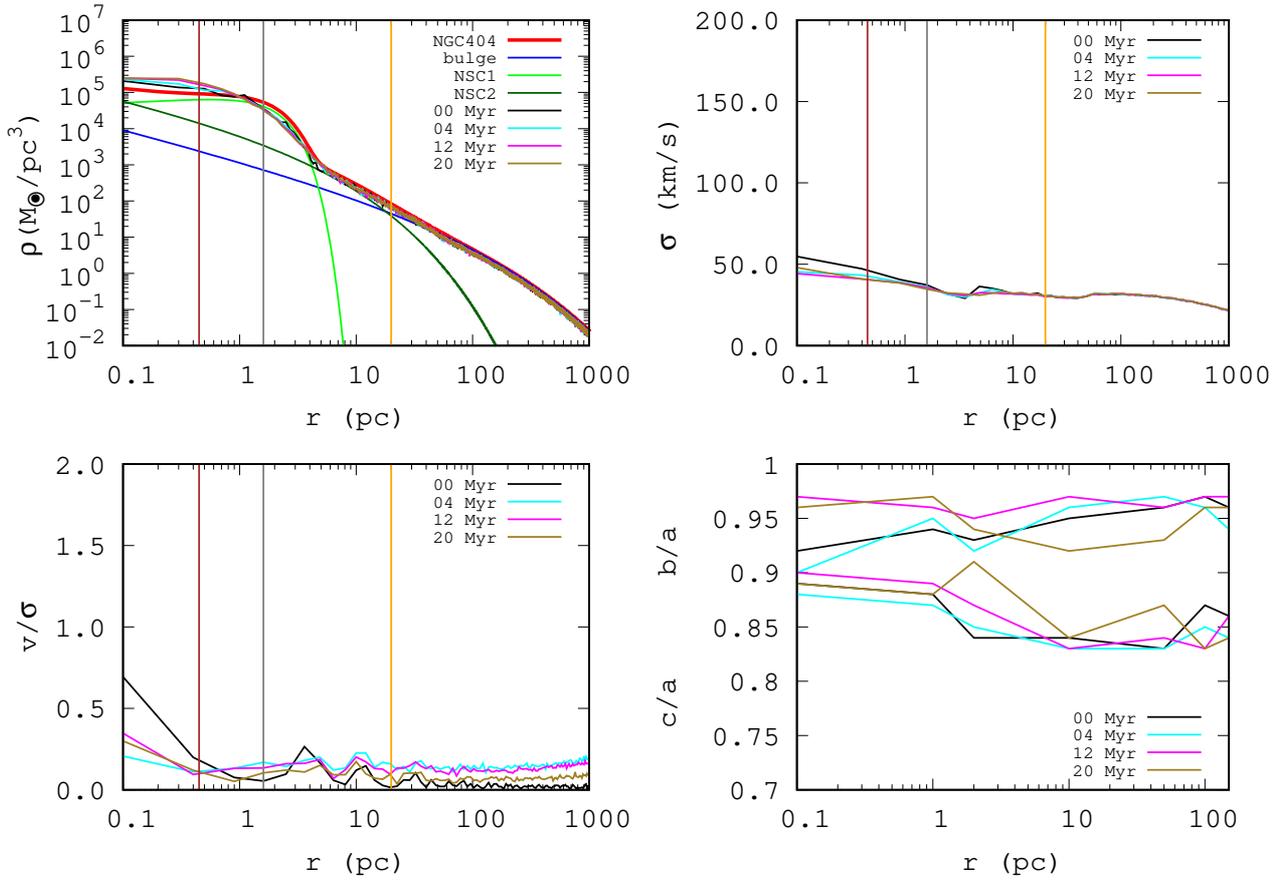}}
		}
		\caption{
			Our initial model and stability analysis for NGC 404. The panels, labels and lines are the same as in figure \ref{fig:ngc5102multi}.  
		} \label{fig:ngc404multi}
	\end{figure*}
	\subsubsection{NGC5206} \label{subsec:ngc5206}
	
	NGC 5206 is a dE/S0 type galaxy with an estimated stellar mass of $2.4 \times 10^9 M_{\odot}$. Its velocity dispersion ranges from $\sigma \sim 35-40$ km/s and the average rotational velocity is about 10 km/s. 
	\citet{ngu18} put an upper mass limit of $4.7 \times 10^5 M_{\odot}$ on a central IMBH with an influence radius of roughly 1 pc. As in the observational decomposition, we represent the stellar distribution of NGC 5206 by two NCSs and a bulge; Table \ref{tab:ngc5206param} gives NSC and bulge parameters for NGC 5206. Note that the central stellar density is $\rho_{\star,c} \simeq 10^{6}$ M$_{\odot}/{\rm pc}^3$ and at the IMBH influence radius it decreases to $\rho_{\star,c} \simeq 10^{5}$ M$_{\odot}/pc^3$. We adopt a mildly triaxial shape ($b/a \sim 0.95$ and $c/a \sim 0.85$) for NGC5206 galaxy model.  
	\begin{table}
		\vspace{-0.5pt}
		\caption{NGC 5206 Galaxy Parameters} 
		\begin{tabular}{c c c c c}
			\hline
			Component & $n$ & $r_{\rm eff}$(pc) & $M_{\star}$ ($10^7$ M$_{\odot}$) & $a_2/a_1$\\
			\hline
			IMBH & $--$ & $1.0$ & $0.047$ & $--$\\
			NSC1 & $0.8$ & $3.4$ & $0.17$ & $0.96$\\
			NSC2 & $2.3$ & $10.5$ & $1.28$ & $0.96$\\
			Bulge & $2.57$ & $986$ & $241$ & $0.98$\\
			\hline
		\end{tabular}\label{tab:ngc5206param}
		\vspace{15pt}
		
		Column~1: Galaxy component. Column~2: Sersic index. Column~3: Effective radius of the galaxy component (note: for the IMBH, we quote the influence radius instead). Column~4: Stellar mass.  Column~5: 2-d axis ratio as measured by GALFIT. \citet{ngu18}.
		\vspace{15pt}
	\end{table}

	Figure \ref{fig:ngc5206multi} presents various key profiles as a function of distance and their stability in our run for the NGC 5206 galaxy model. Again, our initial model reproduces these key parameters reasonably well and they are fairly stable during the course of our stability run.

	\subsubsection{NGC 404} \label{subsec:404}
	
	NGC 404 is S0 type galaxy which is considered to possess a central IMBH and two NSCs \citep{seth10,ngu17}. It has a mean velocity dispersion of about 35 km/s which rises to 40 km/s towards the center, hinting at the presence of an IMBH. The galaxy's mean rotational velocity is only about 8 km/s resulting in a low $v/\sigma \sim 0.2 ~$. \citet{ngu17} found the mass of the central IMBH to be $7 \times 10^4 M_{\odot}$ having an influence radius of about 0.35 pc. The total stellar mass is reported to be $8.6 \times 10^8 M_{\odot}$. Some key parameters of NSCs and bulge reported in \citet{ngu17} are presented in table \ref{tab:ngc404param}.  Shape parameters for this galaxy model are  $b/a \sim 0.95$  and $c/a \sim 0.85$ as well.
	
	\begin{table}
		\vspace{-0.5pt}
		\caption{NGC 404 Galaxy Parameters} 
		\begin{tabular}{c c c c c}
			\hline
			Component & $n$ & $r_{\rm eff}$(pc) & $M_{\star}$ ($10^7$ M$_{\odot}$) & $a_2/a_1$\\
			\hline
			IMBH & $--$ & $0.35$ & $0.007$ & $--$\\
			NSC1 & $0.5$ & $1.6$ & $0.34$ & $0.97$\\
			NSC2 & $1.96$ & $20$ & $1.1$ & $0.95$\\
			Bulge & $2.50$ & $675$ & $84.4$ & $0.99$\\
			\hline
		\end{tabular}\label{tab:ngc404param}
		\vspace{15pt}
		
		Column~1: Galaxy component. Column~2: Sersic index. Column~3: Effective radius of the galaxy component (note: for the IMBH, we quote the influence radius instead). Column~4: Stellar mass.  Column~5: 2-d axis ratio as measured by GALFIT.
		
		\vspace{15pt}
	\end{table}
	
	Figure \ref{fig:ngc404multi} shows our initial model and its stability for 20 Myr. The central density of about $\rho_{\star,c} \simeq 10^{5}$ M$_{\odot}/{\rm pc}^3$ and the IMBH influence radius in our model is about 0.45 pc, making this a close NGC 404 analog. We also witness a low value of $v/\sigma \sim 0.1$.

	\subsubsection{NGC 205} \label{subsec:ngc205}
	
	NGC 205, a nucleated dwarf galaxy, is thought to host an IMBH with an estimated mass ($2.0-5.0 \times 10^4$ M$_{\odot}$) and influence radius $r_{\rm infl.} = 0.14$ parsec~\citep{val05,ngu18}. With such a small IMBH, the contrast between the IMBH and stellar particle mass is 1:80, and we worried that the mass contrast was too similar to accurately model stellar-IMBH scattering. We therefore doubled the particle number to 4 million to achieve a similar mass contrast to our prior work.
	NGC 205 is shown to exhibit no rotation and a flat velocity dispersion of 20 km/s. The total photometric mass estimated for the galaxy is roughly $\sim 10^9$ M$_{\odot}$ \citep{ngu18}. In table 5, we list some key parameters of NGC 205 (table \ref{tab:ngc205param}) as reported by \citep{ngu18}. We choose $b/a \sim 0.95$ and $c/a \sim 0.85$ as shape parameters for this model.
	
	\begin{table}
		\vspace{-0.5pt}
		\caption{NGC 205 Galaxy Parameters} 
		\begin{tabular}{c c c c c}
			\hline
			Component & $n$ & $r_{\rm eff}$(pc) & $M_{\star}$ ($10^7$ M$_{\odot}$) & $a_2/a_1$\\
			\hline
			IMBH & $--$ & $0.14$ & $0.004$ & $--$\\
			NSC & $1.6$ & $1.3$ & $0.18$ & $0.95$\\
			Bulge & $1.4$ & $516$ & $97.2$ & $0.90$\\
			\hline
		\end{tabular}
		
		Column~1: Galaxy component. Column~2: Sersic index. Column~3: Effective radius of the galaxy component (note: for the IMBH, we quote the influence radius instead). Column~4: Stellar mass.  Column~5: 2-d axis ratio as measured by GALFIT.
		
		\vspace{15pt}
	\end{table}\label{tab:ngc205param}

	\begin{figure*}
		
		\centerline{
			\resizebox{0.95\hsize}{!}{\includegraphics[angle=270]{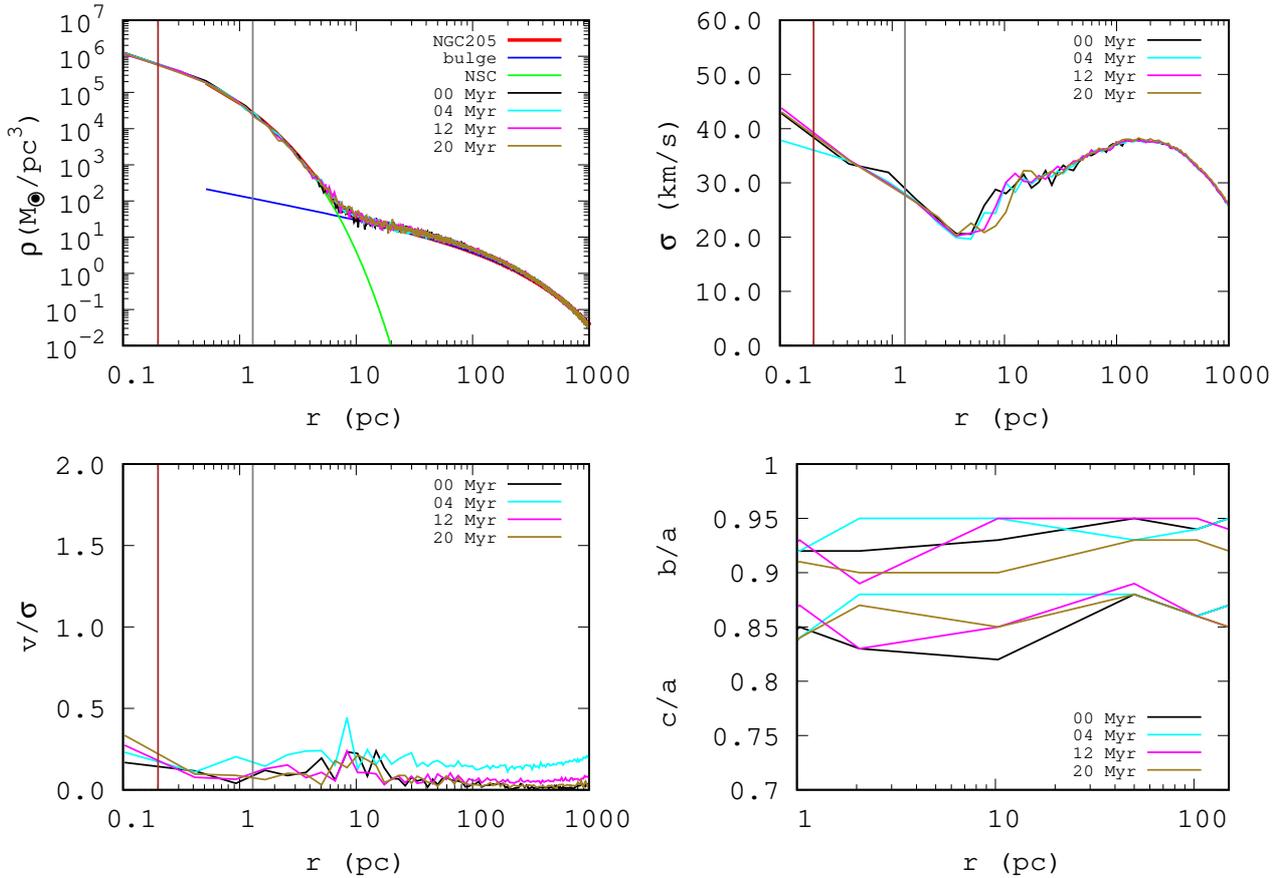}}
		}
		\caption{ The initial model and stability analysis for NGC 205. The panels, labels, and lines are the same as in figure \ref{fig:m32multi}.
		} \label{fig:ngc205multi}
	\end{figure*}
	
	Figure \ref{fig:ngc205multi} shows various parameters of our generated model and also their evolution in stability run that we performed with $\phi$-GPU. The central stellar density is approximately $\rho_{\star,c} \simeq 10^{6}$ M$_{\odot}/{\rm pc}^3$. We find that our model is reasonably stable for the duration of our stability run with properties that mirror those of NGC 205 described in the previous paragraph. 
	\begin{figure}
		\centerline{
			\resizebox{0.95\hsize}{!}{\includegraphics[angle=270]{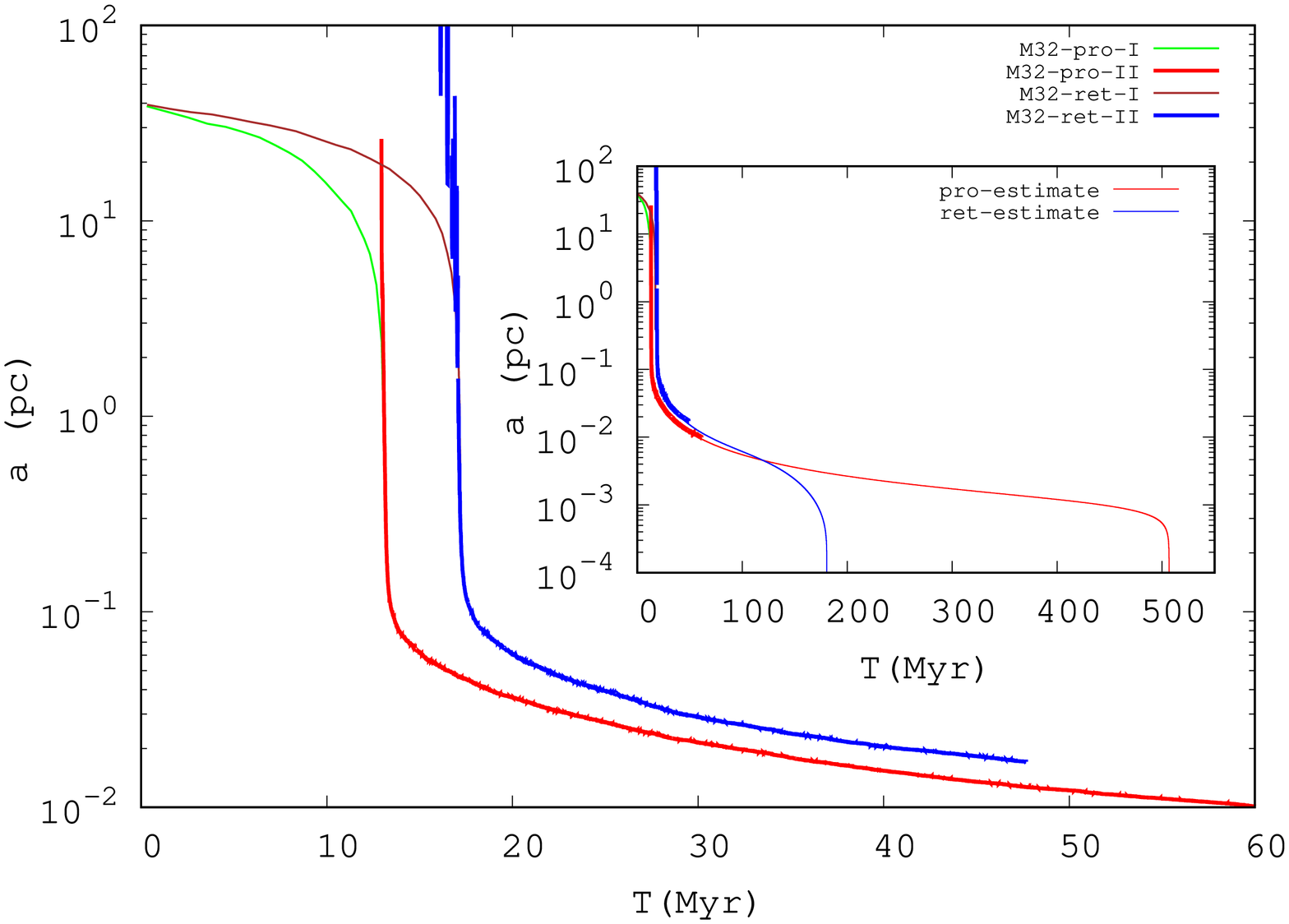}}
		}
		\centerline{
			\resizebox{0.95\hsize}{!}{\includegraphics[angle=270]{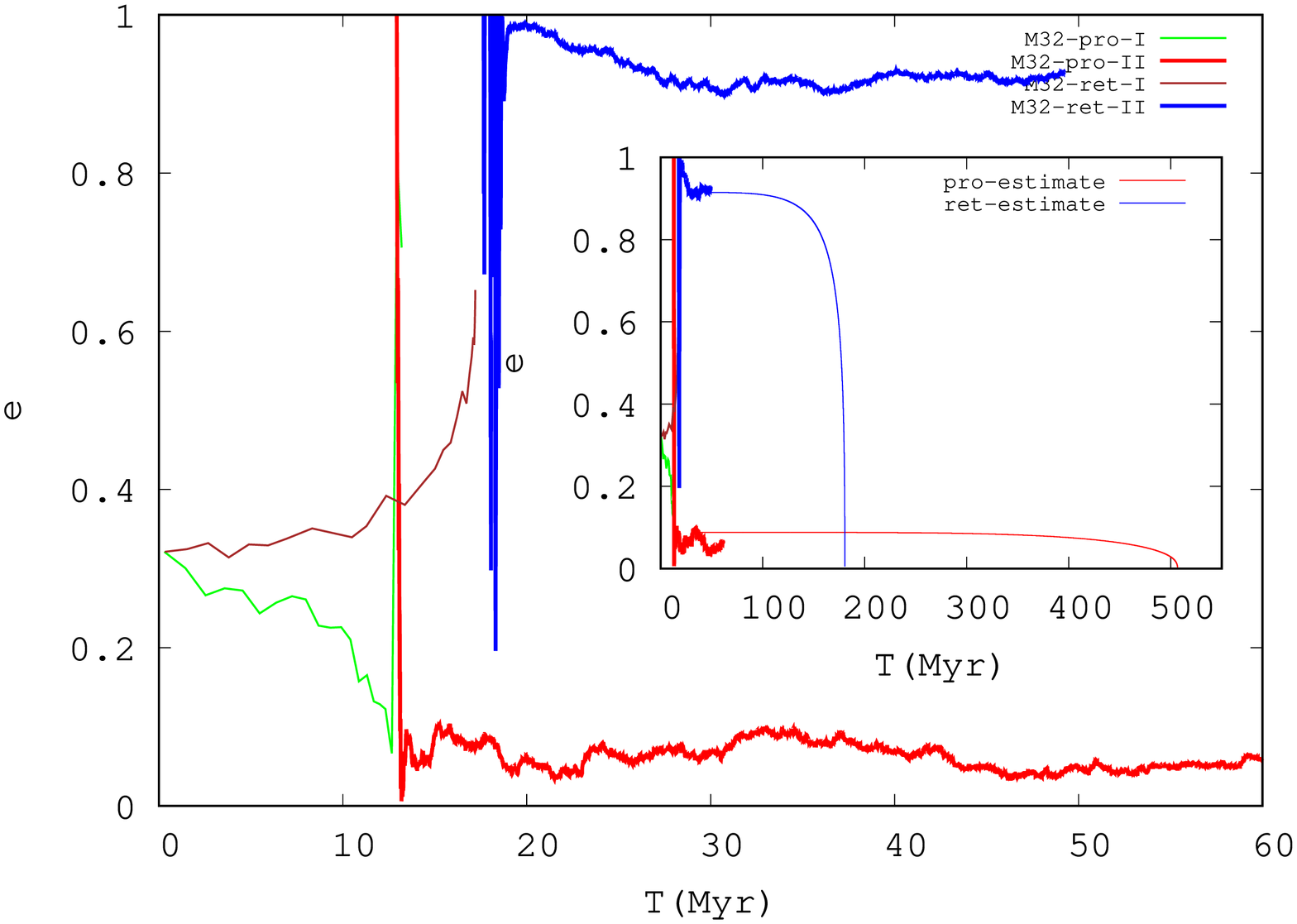}}
		}
		\caption{
			Time evolution of the IMBH binary semi-major axis (top panel) and eccentricity (bottom panel) for both pro- and retrograde models of M32. Inset panels show the complete evolution of the semi-major axis and eccentricity for an IMBH pair from the start of the run to merger, as described in section \ref{subsec:reciepe}. In Phase I, the IMBH system is unbound, so the semimajor axis and eccentricity is determined by the apo and pericenter passes of the inspiral. Once the system is bound (Phase II, respresented by thicker lines), we use the standard Keplerian definitions. Once the simulation ends, we estimate subsequent evolution (see text), represented by thin lines in the inset plots. 
		} \label{fig:m32param}
	\end{figure}
	\begin{figure}
		\centerline{
			\resizebox{0.95\hsize}{!}{\includegraphics[angle=270]{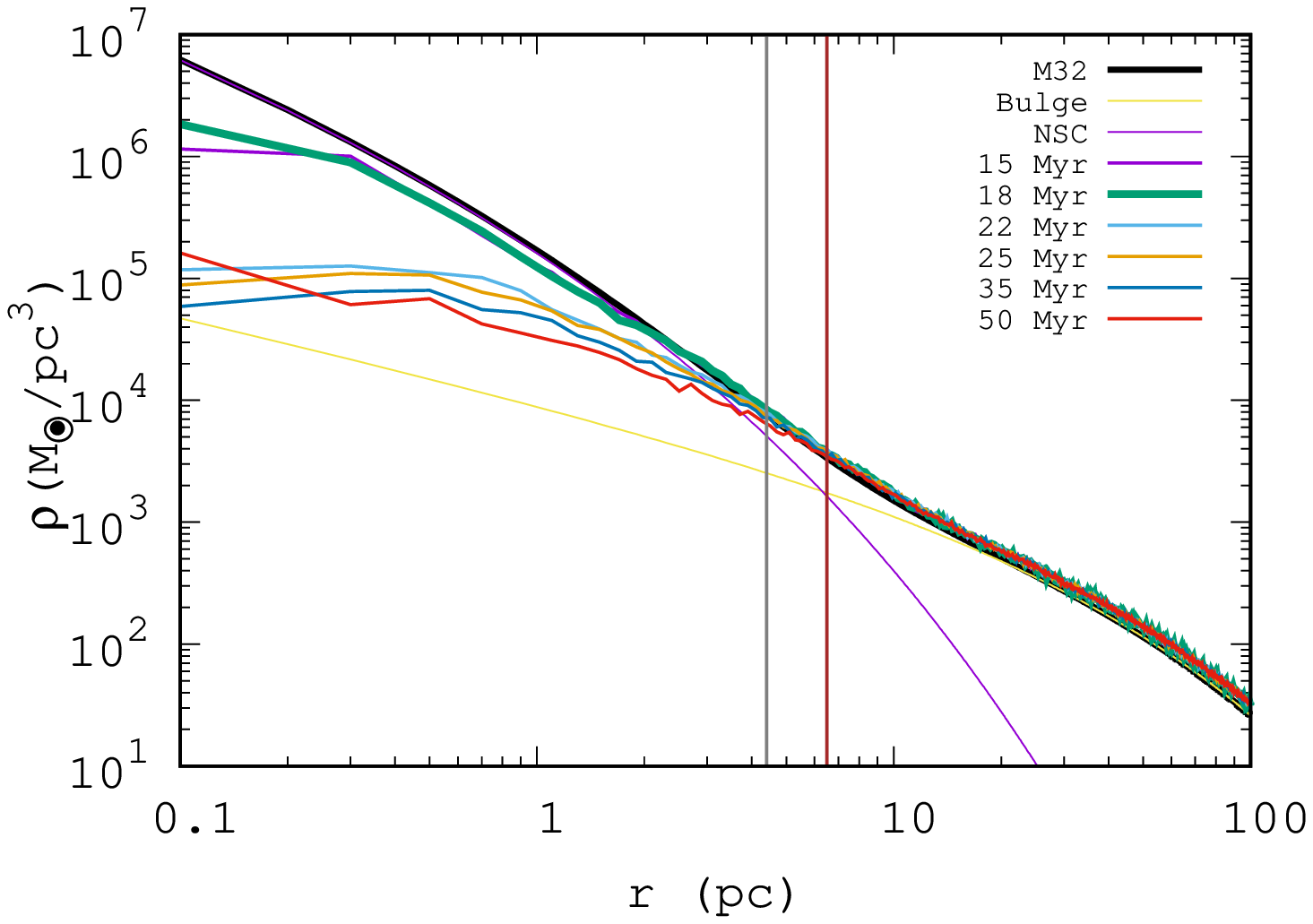}}
		}
		\centerline{
			\resizebox{0.95\hsize}{!}{\includegraphics[angle=270]{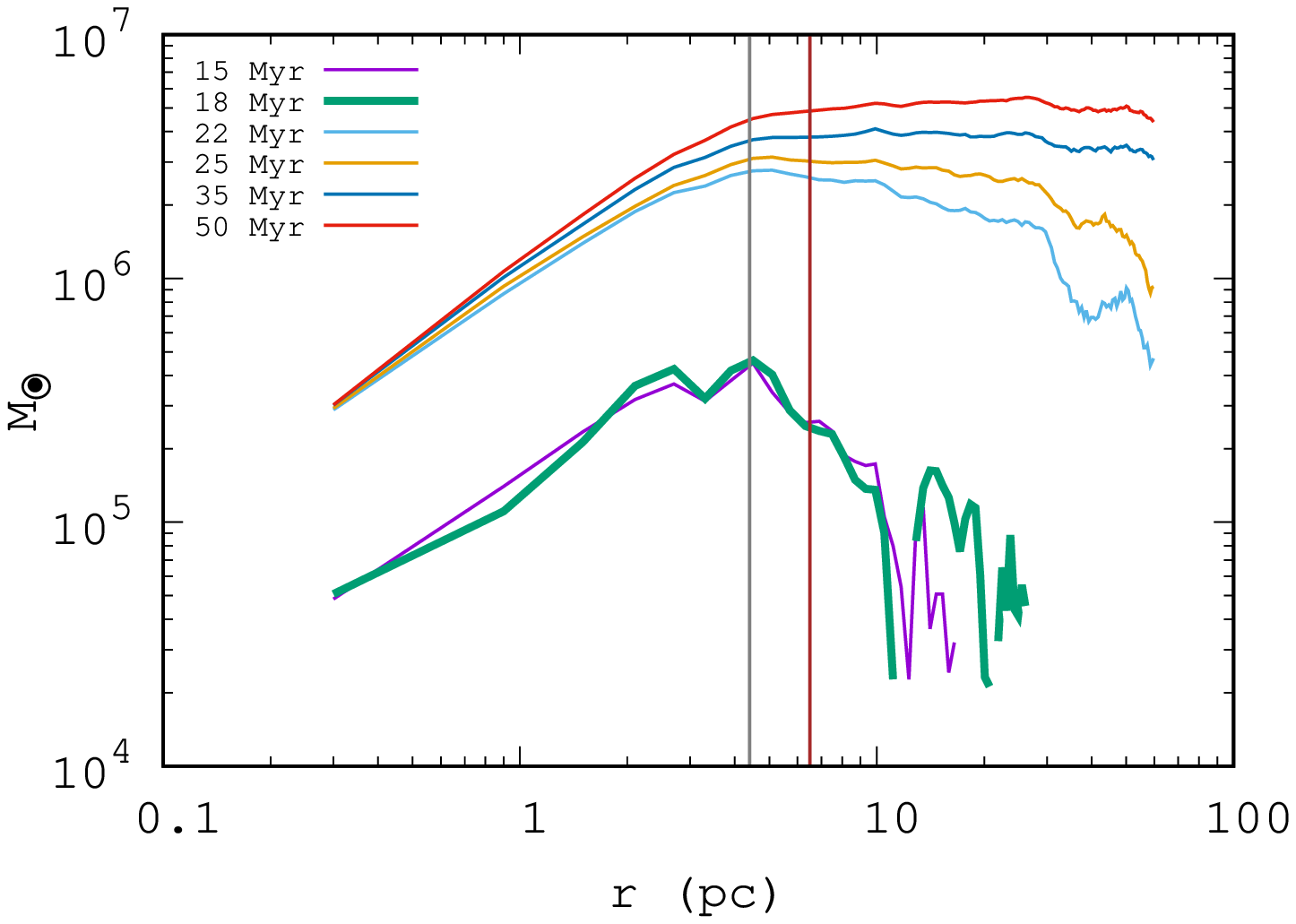}}
		}
		\caption{
			Time evolution of the density profile (top) and cumulative mass deficits (bottom) for the retrograde M32 run. The thick green curve marks the binary formation time. The vertical brown line denotes the influence radius of the total binary after hard IMBH binary formation. The grey vertical line is the NSC effective radius.
		} \label{fig:m32dm}
	\end{figure}
	\subsection{Initial orbits} \label{subsec:iniorb}
	
	As we generate our models in equilibrium with an IMBH, our primary black hole is initially at rest at the origin within the stellar cusp. We introduce a secondary IMBH at a distance of roughly $50$ pc with $50\%$ of the circular velocity for the model, resulting in an intermediate initial eccentricity of $0.5-0.6$. For M32 and NGC 5102 galaxies, which exhibit significant rotation, we choose two orbital configurations of the secondary IMBH orbit; prograde and retrograde with respect to galaxy's angular momentum. In the unbound phase (Phase I in the figures), the orbital parameters of the secondary IMBH are calculated from apo- pericenters of the inspiral, and in the bound state (Phase II in the figures), they are calculated using Keplerian equations. 
	
	\subsection{Binary evolution} \label{subsec:reciepe}
	
	We do not follow IMBH binary evolution until the final coalescence, due to extremely high computational requirements of post-Newtonian simulations with a large particle number, greater than 2 million in our case, and the further need for numerical relativity once the IMBHs enter the strong gravity regime. Instead, we evolve the IMBH binary through the $3-$body scattering phase and into the weak gravitational wave phase with a recipe we previously developed and shown to match well with post-Newtonian simulations \citep{khan+12b,khan18b}. In the transition to the gravitational wave regime, the semi-major axis of IMBH binary evolves due to combined effect of stellar and gravitational wave hardening. We adopt the stellar hardening rates and eccentricity from the late phase of our direct N-body simulations and add gravitational wave hardening from \citet{peters+63} until merger. 
	
	\section{Results}\label{sec:results}
	

	\begin{figure}
		\centerline{
			\resizebox{0.95\hsize}{!}{\includegraphics[angle=270]{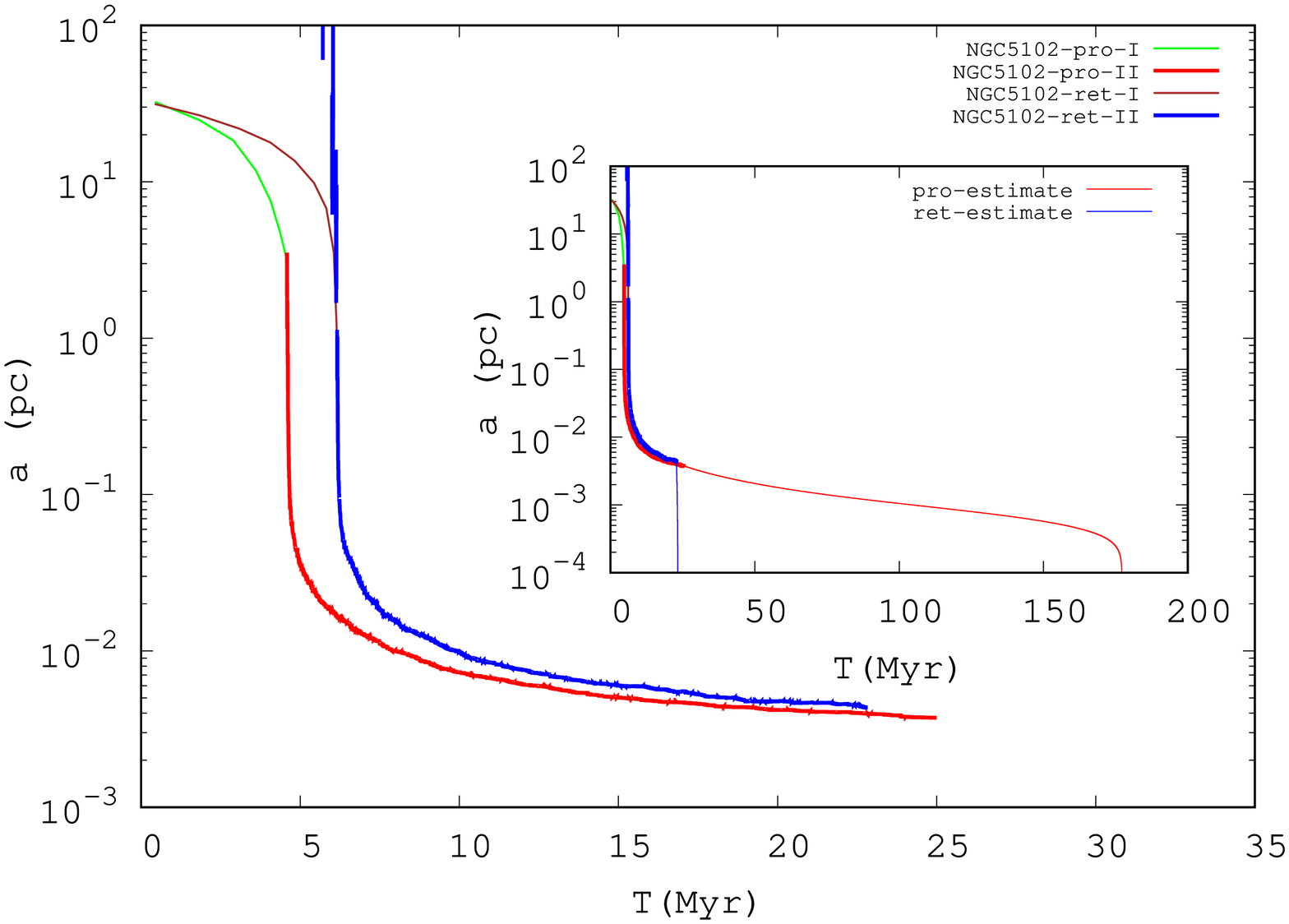}}
		}
		\centerline{
			\resizebox{0.95\hsize}{!}{\includegraphics[angle=270]{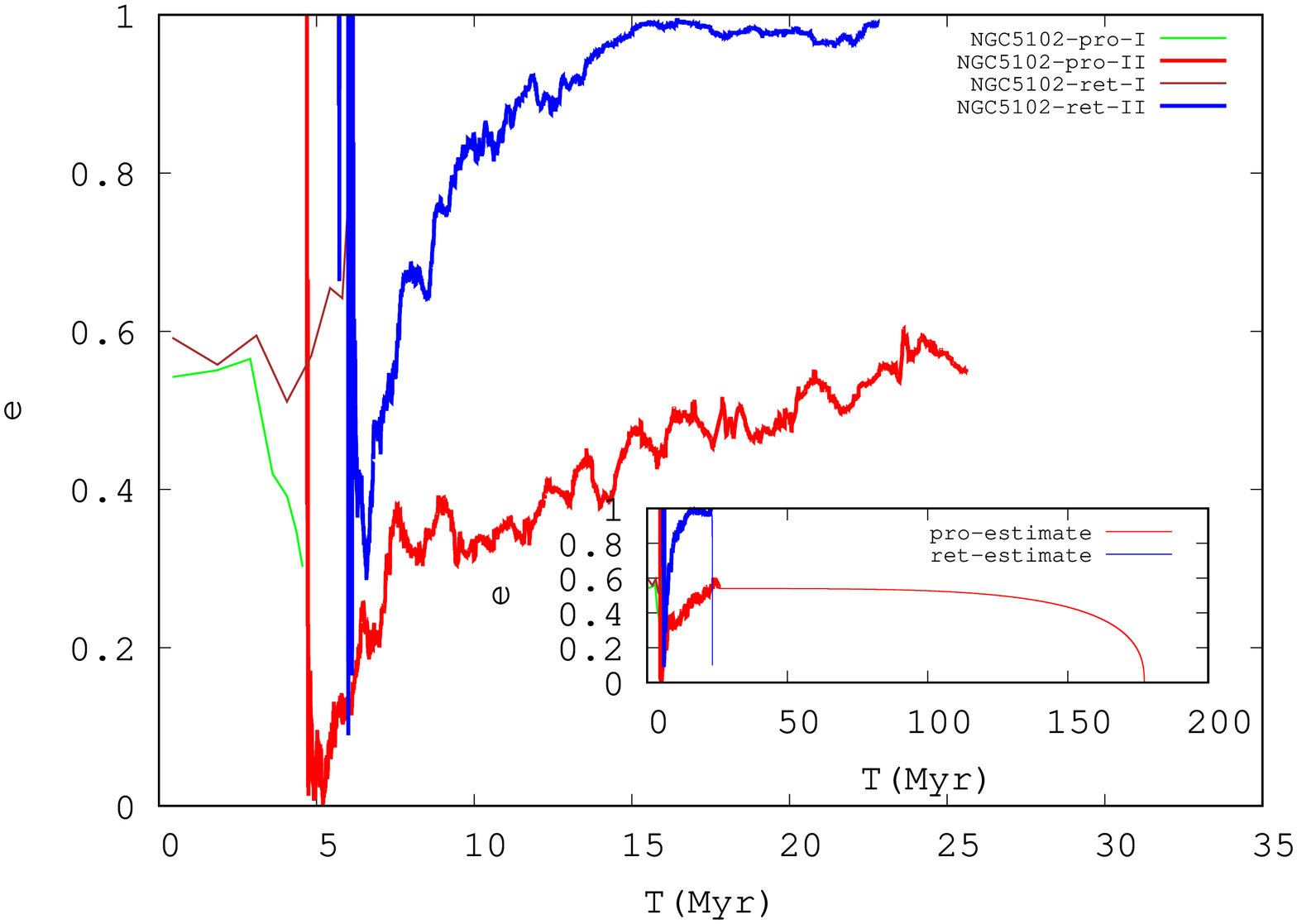}}
		}
		\caption{
			IMBH pair evolution for NGC 5102, as explained in figure \ref{fig:m32param}. 
		} \label{fig:ngc5102param}
	\end{figure}

	\begin{figure}
		\centerline{
			\resizebox{0.95\hsize}{!}{\includegraphics[angle=270]{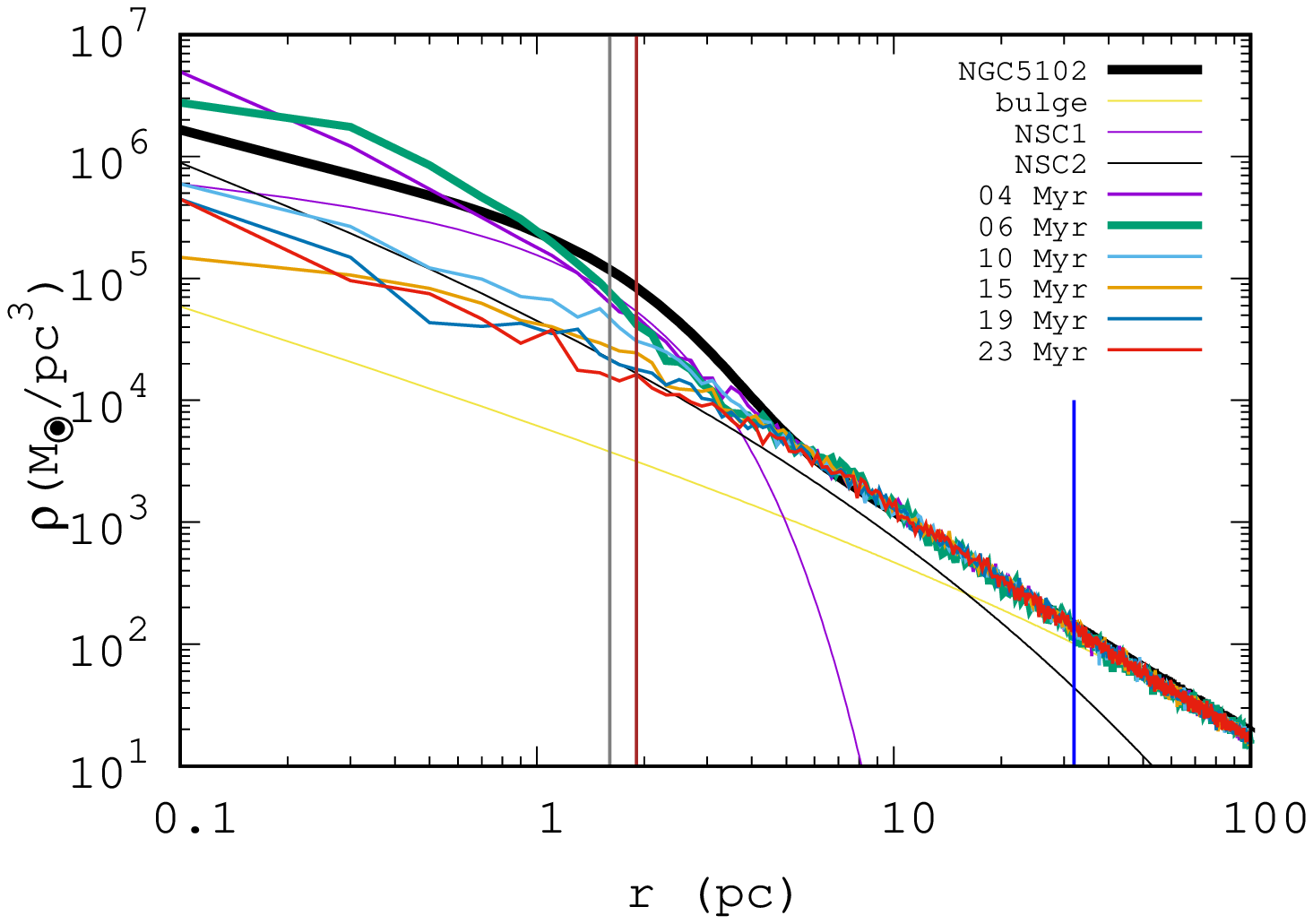}}
		}
		\centerline{
			\resizebox{0.95\hsize}{!}{\includegraphics[angle=270]{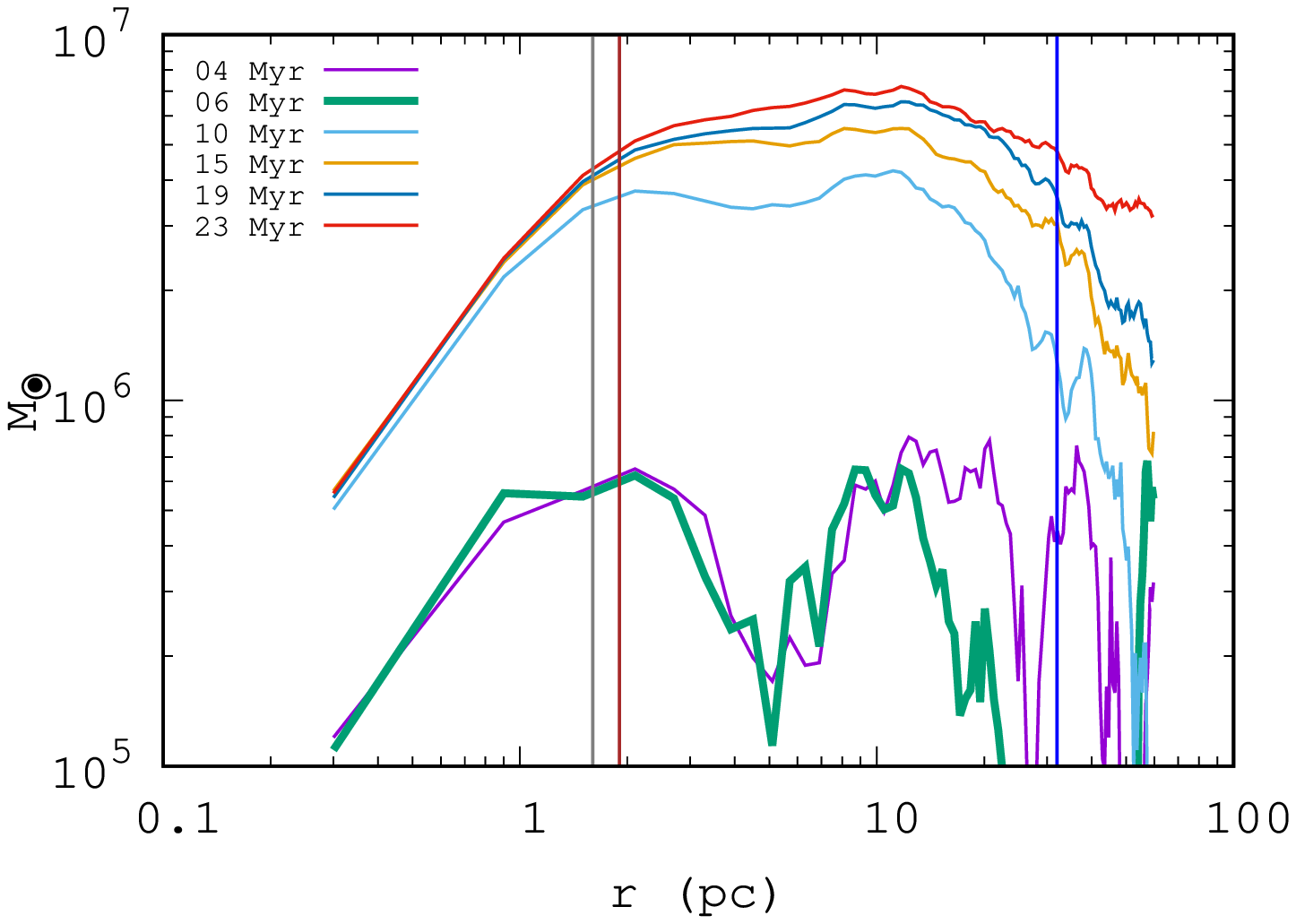}}
		}
		\caption{
			Density profiles (top) and cumulative mass deficits (bottom) for the NGC 5102 run at various times in Myr. The thick green curve marks the IMBH binary formation time. As earlier, the vertical brown line is the IMBH binary's influence radius. Vertical grey and blue lines are the effective radii of NSC1 and NSC2, respectively.
		} \label{fig:5102dm}
	\end{figure}
	
	Here we present results of our study of IMBH pair evolution in nucleated dwarfs.

	\subsection{M32}\label{subsec:resultsm32}

	To study IMBH pair evolution in the M32 nucleus, we introduce a secondary black hole with 1/5th of the primary mass ($=5 \times 10^{5} M_{\odot}$) at a distance of $r_{init}$ = 54 pc -- half of the bulge effective radius. The secondary IMBH is given 50 $\%$ of circular velocity at $r_{\rm init}$. We set the secondary on pro- and retrograde orbits with galaxy's sense of rotation.
	Figure \ref{fig:m32param} shows the time evolution of the semi-major axis and eccentricity of the IMBH pair orbit. In the dynamical friction phase, the prograde orbit shrinks efficiently, while the retrograde orbit experiences a delay (top-left panel), evident from the longer binary formation time for the retrograde case: binary formation happens at $T_{\rm b} = 13$ Myr for the prograde orbit and $T_{\rm b} = 18$ Myr for the retrograde orbit. The prograde binary forms with a very low eccentricity and stays almost circular throughout its evolution, whereas the retrograde binary forms with $e \sim 0.99$ and reduces to $e \sim 0.9$ in hard binary phase,  growing slowly again with time. We also observe that the binary orbital plane {\it flips} in the retrograde case so that it aligns with the system angular momentum, as witnessed in \citet{kha20}.  
	These results are in broad agreement with our earlier studies \citep{holley+15,kha20}, where we found rapid orbital decay and circularization for co-rotating binaries and delayed orbital decay accompanied by eccentricity growth for counter-rotating orbits over a wide range of mass ratios. 
	
	In both cases, at the end of the dynamical friction phase, Keplerian binaries form and the semi-major axes shrink rapidly.  The rapid phase of binary shrinking transforms to a more steady one in the hard binary regime. For the prograde orbit, the hard binary transition happens at $T_{\rm hb} = 15$ Myr and for retrograde case it takes place at $T_{\rm hb} = 20$ Myrs. The decrease in shrinking rate in this phase is directly linked with the decrease in central density as the binary scours the nucleus by $3-$body scattering encounters (see top panel of figure \ref{fig:m32dm}).  
	
	
	The hardening rate $`s`$ obtained by finding slope of a straight line fit to $1/a$ evolution for prograde and retrograde binaries are very similar,  $1.94~pc^{-1}$/Myr and $1.77~pc^{-1}$/Myr, respectively. Subsequent evolution (as described in section \ref{subsec:reciepe}) of the IMBH binary into the gravitational wave phase followed until the estimated coalescence of IMBHs is presented in the inset panels figure \ref{fig:m32param}. It shows the semi-major axis and eccentricity evolution of the IMBH binary for the prograde and retrograde runs. 
	The merger time for black holes in the prograde configuration is 510 Myr. In the retrograde configuration, we estimate merger times for eccentricity value of $e \sim 0.9$. A retrograde binary results in a much faster black hole merger time of 180 Myr due to its high eccentricity. Even with this unequal mass binary, we find merger times of the order of few hundred Myrs thanks to the higher central stellar density in M32's NSC.              
	

	The evolution of the binary IMBH partially disrupts the NSC, particularly as the binary transitions from the soft to hard binary phase. We note a clear depletion of stellar mass that extends beyond the effective radius of the NSC, suggesting that stellar encounters originating from {\it bulge} are also crucial for binary IMBH hardening. The influence radius of the binary IMBH is extended from 1.6 pc to 2.9 pc,  mainly due to core scouring, but also because the secondary IMBH must be taken into account. 
	
	The bottom panel of figure \ref{fig:m32dm} shows cumulative mass deficits, calculated by comparing mass profiles at an initial and later specific time. There is a clear abrupt jump in the mass deficit between the time of binary formation and the hard binary phase. Later, mass deficits evolve steadily until the end of our run.  The mass deficits peaks at 26.7 pc, well outside the NSC effective radius (4.4 pc), further strengthening our inference that in the hard binary phase, stellar encounters with the binary are drawn from the centrophilic orbits in the bulge at larger radii. We estimate total mass deficit induced by binary to be $1.85 \times M_{\rm IMBHs}$.

	\subsection{NGC 5102}\label{subsec:5102param}
	
	NGC 5102's rotation is mild at the center and increases with radius (figure \ref{fig:ngc5102multi}). We launch an equal mass secondary black hole on pro- and retrograde orbits at a distance of 50 pc from the center with a total initial velocity that is 50 $\%$ of the circular velocity. IMBH pair evolution is shown in figure \ref{fig:ngc5102param}. The semi-major axis of the secondary IMBH shrinks down below 0.1 pc in roughly 5 Myrs at which point they form a binary with the primary IMBH. As before, however, the binary formation time is earlier in the prograde configuration. The IMBH binary semi-major axis evolution is very similar for both the scenarios, as is evident the nearly equal hardening rates -- $s = 8.8 {\rm pc}^{-1}$/Myr for the prograde and $s = 8.3 {\rm pc}^{-1}$/Myr for the retrograde orientation, though again the prograde binary shrinks slightly faster. For the prograde case, the binary forms with a low eccentricity and increases gradually to an intermediate value of $e \sim 0.54$ when we stop our simulation (figure \ref{fig:ngc5102param}). This result is somewhat contrary to the M32 case (figure \ref{fig:m32param}), as well as our earlier studies \citep{holley+15,kha20} where we find that on prograde binaries retain low eccentricities in the hard binary regime. A key difference, however, is that the NGC5102 nucleus is only very mildly rotating, so the effect of host rotation on the orientation of the binary orbit is much less significant. Here, IMBH coalescence happens in 178 Myr if we assume a constant eccentricity as the binary transitions from the stellar dynamical to gravitational wave phase. Since the eccentricity is growing when we stop the run, this merger time can be taken as an upper limit.
	
	In the retrograde orbit case, the IMBH binary forms with an intermediate value $e \sim 0.6$ and after a brief period of circularization, it grows to very high eccentricity of $e \sim 0.99$ in the hard binary phase. \citet{kha20} showed that retrograde black hole binary orbits are vulnerable to flipping its orbital plane to a prograde configuration, but, if the black hole binary can {\it maintain} a retrograde orbit while entering the hard binary regime, the eccentricity can grow to unity. In the case of NGC 5102, the relatively mild central rotation is not enough to flip the binary's plane, so the binary retains its retrograde orientation and grows highly eccentric. This high eccentricity ensures that the IMBHs merge very efficiently on nearly radial orbits in about 23.42 Myr. In fact, despite the inevitable circularization of the binary orbit once gravitational radiation is strong, we estimate that the binary can enter the LISA regime with a detectable eccentricity of order 0.01~\citep{peters+63}.  Since this level of eccentricity may be observable with LISA, we plan to revisit the eccentricity evolution with more detailed post-Newtonian simulations in the future.

	Figure \ref{fig:5102dm} shows the evolution of density (top) and cumulative mass deficits (bottom) during our counter-rotating run. We see that the central density cusp drastically flattens in the loose binary phase and keeps on evolving due to stellar ejection in the hard binary regime; indeed the inner nuclear star cluster is entirely disrupted.  The binary's sphere of influence has expanded to 1.9 pc; this is outside the initial effective radius of NSC1, but still roughly 10 times smaller than the effective radius of NSC2. The mass deficit ($4.1 \times M_{IMBHs}$) peaks at 11.7 pc, well inside effective radius of NSC2 suggesting that stellar encounters in hard binary phase originate inside the NSCs and bulge does not play a significant role in the binary's hardening.

	\subsection{NGC5206} \label{subsec:5206param}
	
\begin{figure}
		\centerline{
			\resizebox{0.95\hsize}{!}{\includegraphics[angle=270]{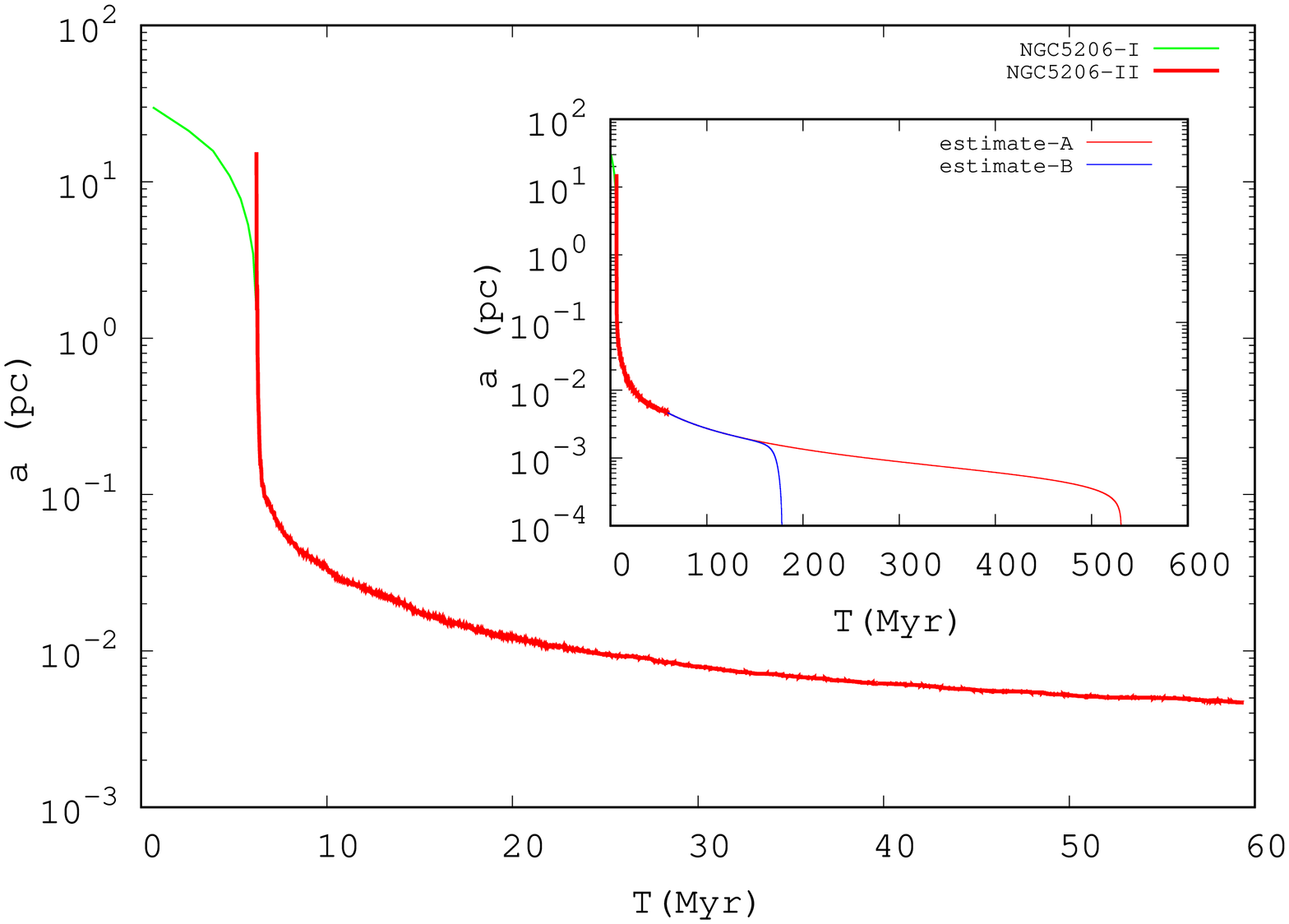}}
		}
		\centerline{
			\resizebox{0.95\hsize}{!}{\includegraphics[angle=270]{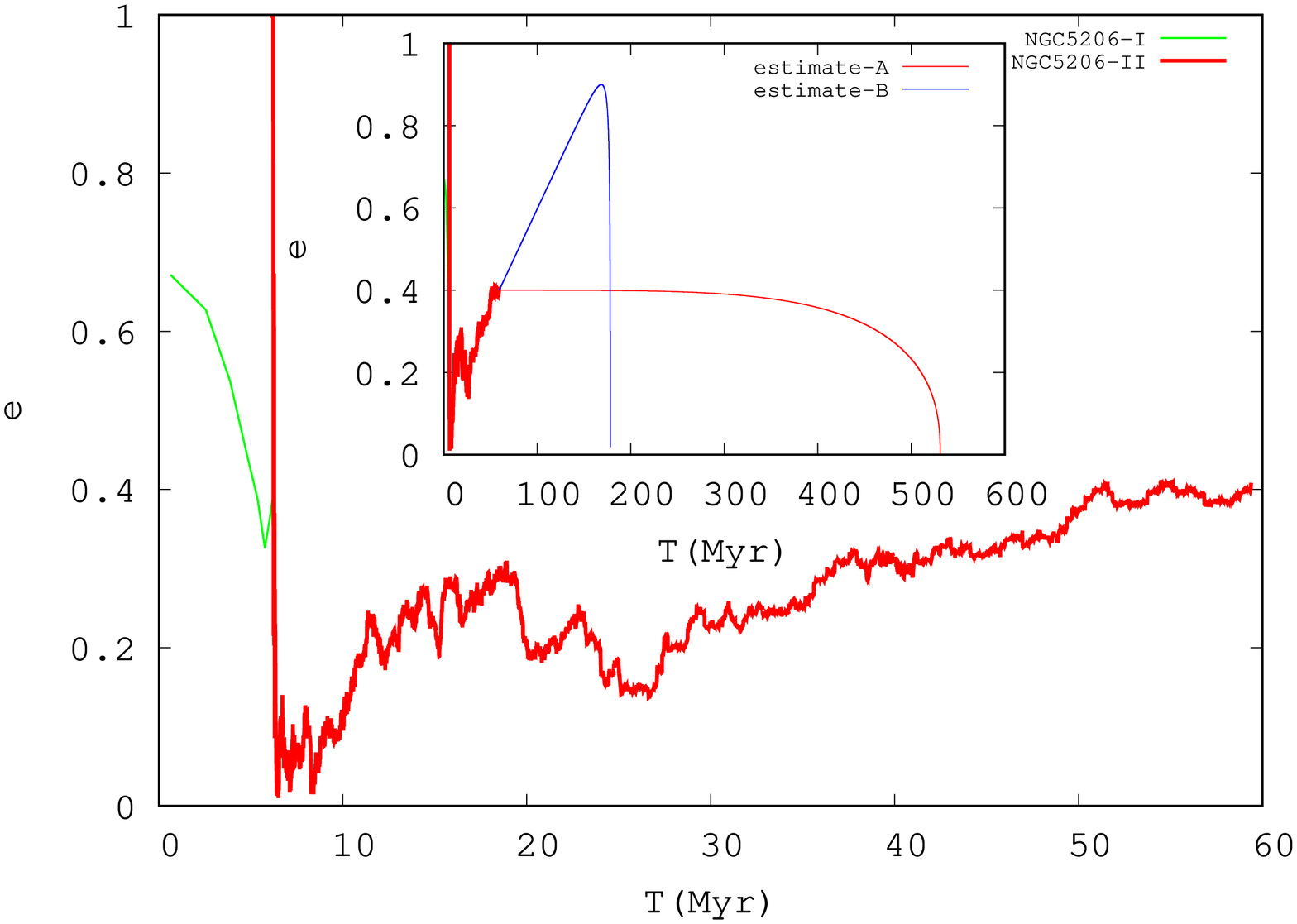}}
		}
		\caption{
			IMBH pair evolution for NGC5206. Inset panels show evolution for a constant eccentricity in the case of `estimate-A` and `estimate-B` is obtained by extrapolating the eccentricity growth. 
		} \label{fig:ngc5206param}
	\end{figure}

	As in NGC 5102, we introduce an equal mass secondary IMBH at a distance of 50 pc from center with $50 \%$ of the circular velocity; since the rotation in this dwarf is negligible, we explore only one orbital configuration. The time evolution of the IMBH pair is depicted in figure \ref{fig:ngc5206param}. Black holes form a binary roughly at $T_b \sim 6$ Myr with very small eccentricity, and a hard IMBH binary forms at around $T_b \sim 8$ Myr. The eccentricity grows slowly in the hard binary phase but is still increasing at the end of our run. The estimated hardening rate of the IMBH binary for this model is $3.73 {\rm pc}^{-1}/Myr$, consistent with expectations from its high density, low rotation, and mild flattening.


	As earlier, IMBH binaries are further evolved using the recipe explained in section \ref{subsec:reciepe}. Complete evolution is shown in inset panels of figure \ref{fig:ngc5206multi} for two different scenarios: `estimate-A` adopts a constant eccentricity $e = 0.4$ observed at the end of simulation, and `estimate-B` extrapolates the eccentricity evolution. For `estimate-A`, the IMBHs merge in 532 Myr. For `estimate-B`, we predict a shorter merger time of 178 Myr as $e$ approaches 0.9.
	
	\begin{figure}
		\centerline{
			\resizebox{0.95\hsize}{!}{\includegraphics[angle=270]{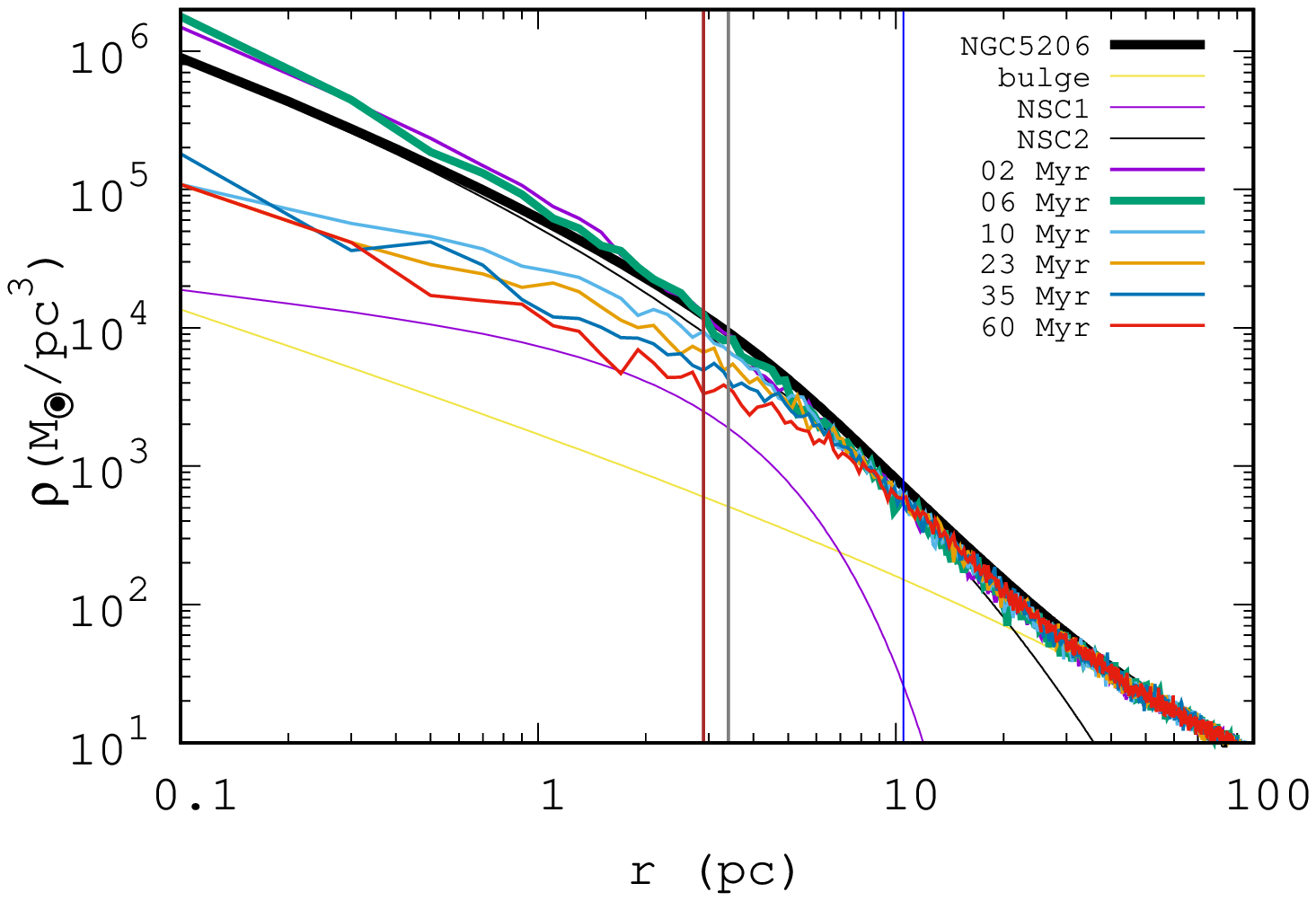}}
		}
		\centerline{
			\resizebox{0.95\hsize}{!}{\includegraphics[angle=270]{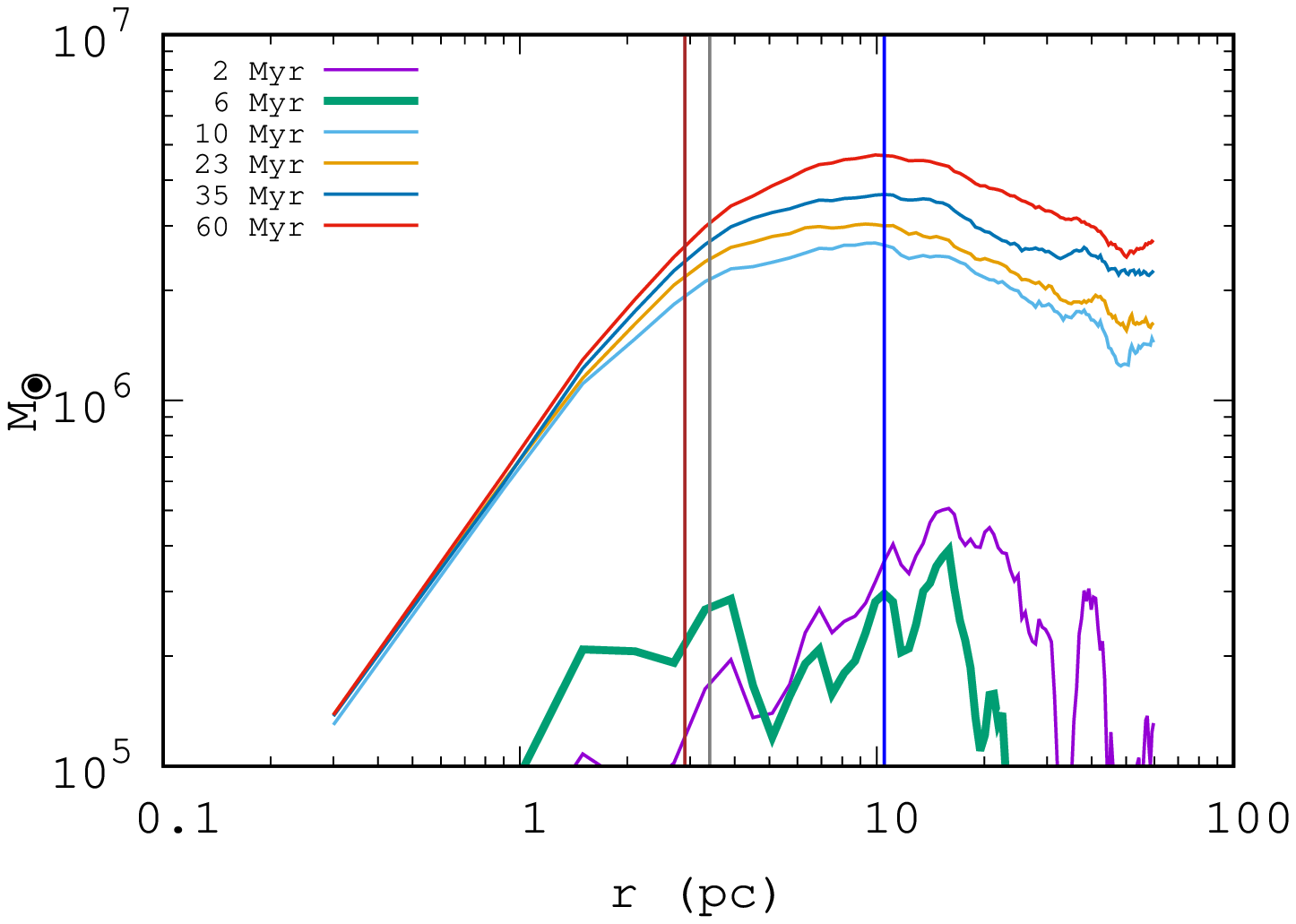}}
		}
		\caption{
			Density profiles (top) and cumulative mass deficits (bottom) for NGC5206 run at various times (in Myr). The thick green curve marks the binary formation time. As earlier, the vertical brown line is the IMBH binary's influence radius. Vertical grey and blue lines are the effective radii of NSC1 and NSC2, respectively.
		} \label{fig:ngc5206dm}
	\end{figure}
	
	Figure \ref{fig:ngc5206dm} shows the evolution of density profiles (top) and mass deficits (bottom) during the NGC 5206 run. The influence radius increases by a factor of 3 to 2.9 pc and a significant flattening of the density profile occurs near the influence radius as NSC1 disrupts. 
	Mass deficits ($4.9 \times M_{IMBHs}$) peak at 9.9 pc, which is close to the effective radius of NSC2. However, the contribution to the stellar density at the influence radius due to the NCSs are still two orders of magnitude greater than that of bulge, suggesting that stellar interactions from NSCs are the primary driver of the IMBH evolution.    
	
	\subsection{NGC404} \label{subsec:404param}
	
\begin{figure}
		\centerline{
			\resizebox{0.95\hsize}{!}{\includegraphics[angle=270]{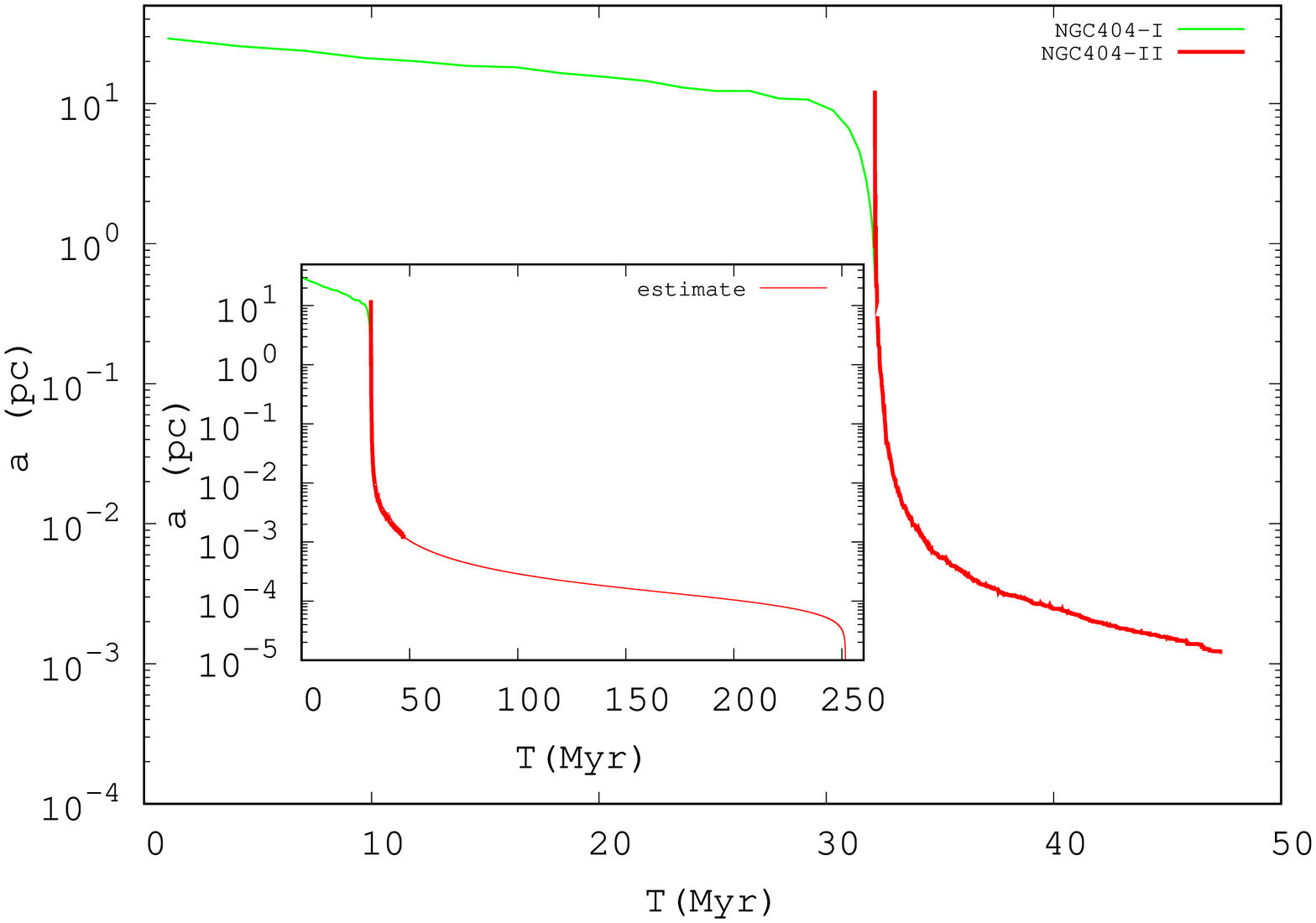}}
		}
		\centerline{
			\resizebox{0.95\hsize}{!}{\includegraphics[angle=270]{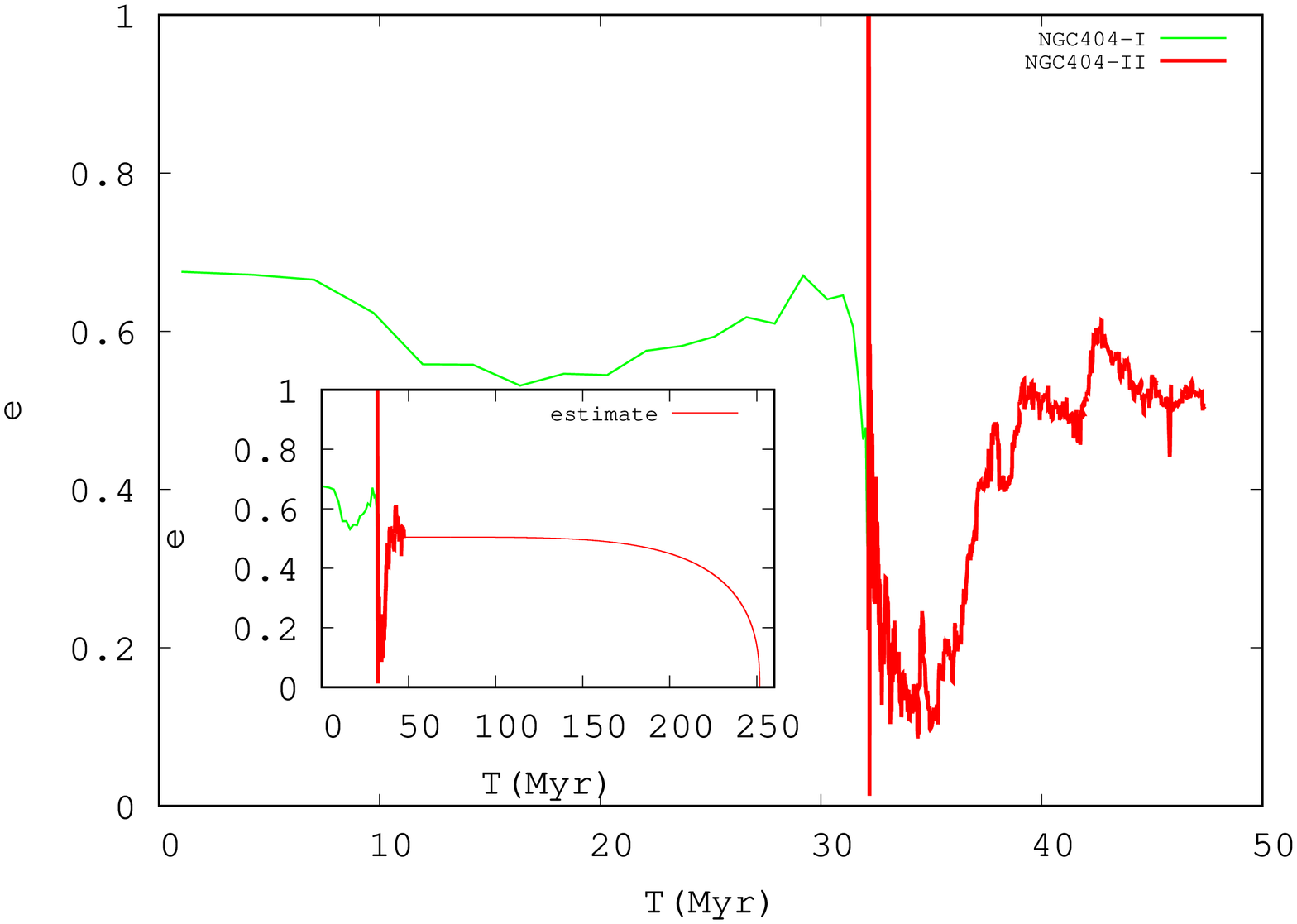}}
		}
		\caption{
			IMBH pair evolution for NGC404. The inset panels show the estimated evolution for a constant eccentricity after the simulation stops.
		} \label{fig:ngc404param}
	\end{figure}

\begin{figure}
		\centerline{
			\resizebox{0.95\hsize}{!}{\includegraphics[angle=270]{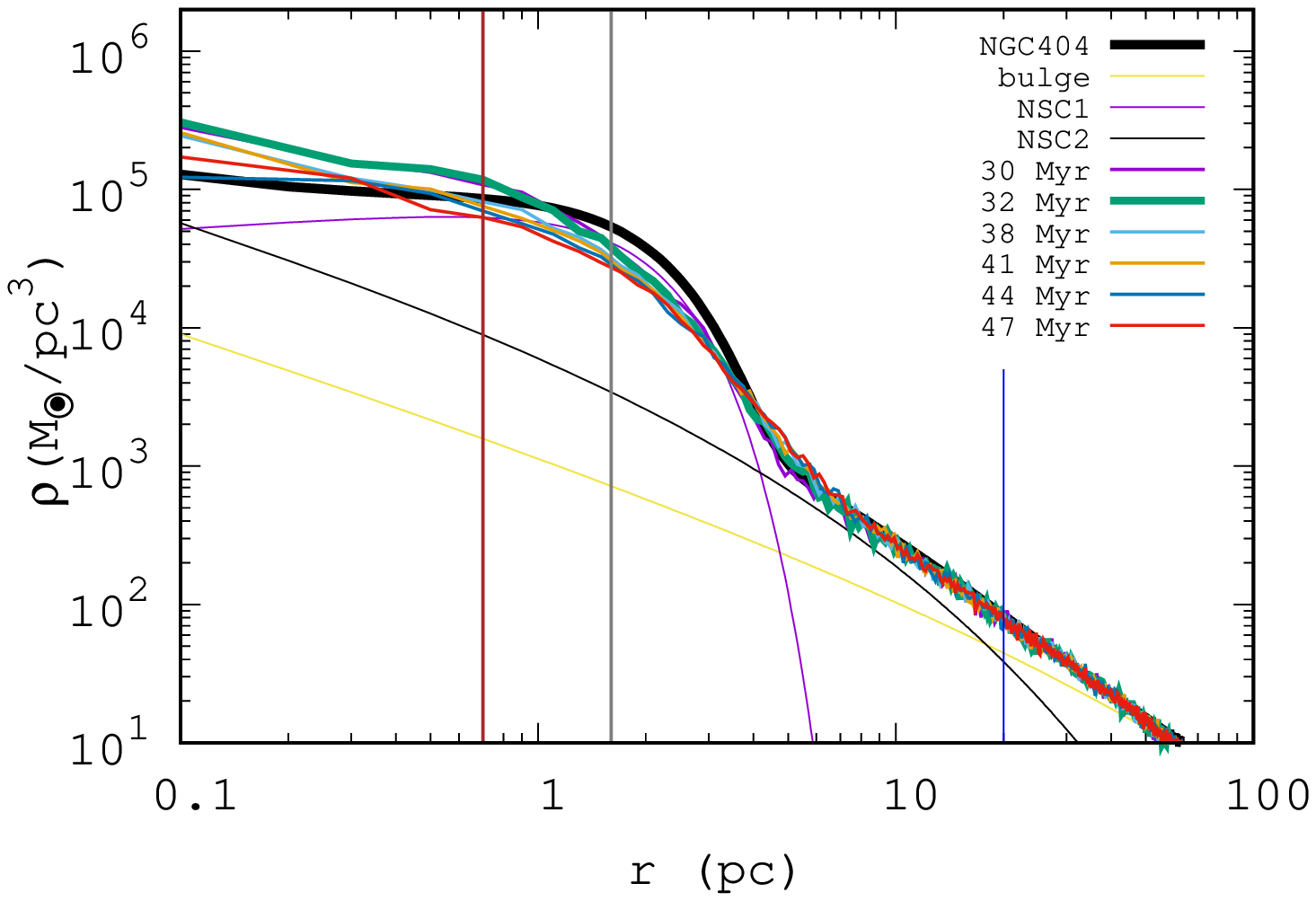}}
		}
		\centerline{
			\resizebox{0.95\hsize}{!}{\includegraphics[angle=270]{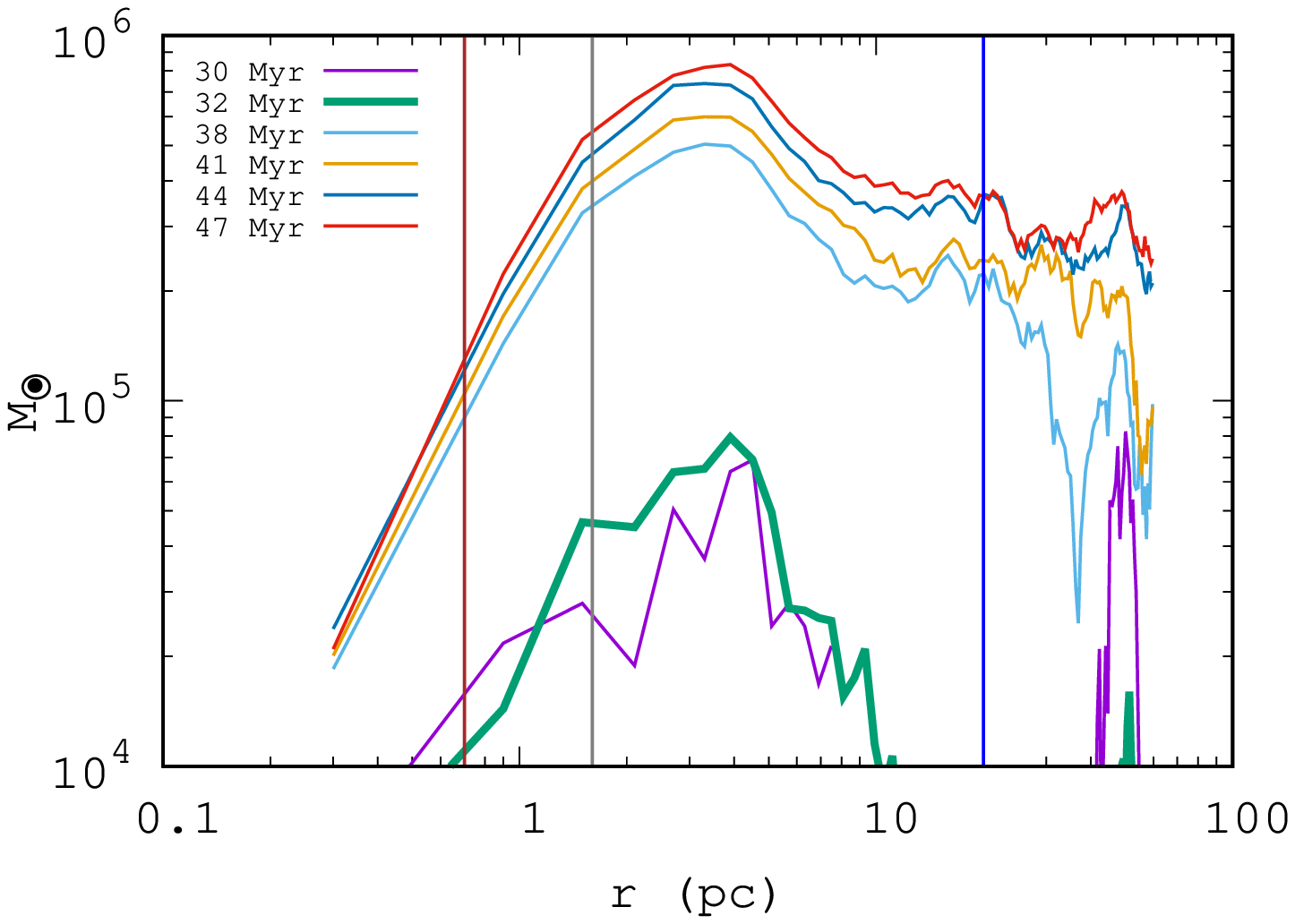}}
		}
		\caption{
			Density profiles (top) and cumulative mass deficits (bottom) for the NGC404 run at various times (in Myr). The thick green curve marks the binary formation time. As earlier, the vertical brown line is the IMBH binary's influence radius. The vertical grey and blue lines are the effective radii of NSC1 and NSC2, respectively.
		} \label{fig:ngc404dm}
	\end{figure}

	As before, we introduce an equal mass ($7 \times 10^4 M_{\odot}$) secondary at 50 pc from the center with a velocity that is $50 \%$ of the circular velocity. An IMBH binary forms with low eccentricity that grows to an intermediate value of $e \sim 0.5$, remaining more or less stable around that value. Figure \ref{fig:ngc404param} shows key orbital elements of the run. IMBH binary hardening rates are particularly high $s = 49~{\rm pc}^{-1}$/Myr. Estimated binary evolution is presented in the inset panels of figure \ref{fig:ngc404param}. The IMBHs merge efficiently in 251.44 Myr thanks to high hardening rates.  The stellar density decreases slightly and remains around (($10^5 M_{\odot}/pc^3$)) between 0.1 - 3 pc (see fig. \ref{fig:ngc404dm}). Although the influence radius of the IMBH binary increased to 0.7 pc due to core scouring, the density at the influence radius did not change much, which allowed the high hardening rate to persist.
	The mass deficits ($5.85 \times M_{IMBHs}$) in figure \ref{fig:ngc404dm} peak at 3.9 pc, well within the effective radius of NSC2. Again, we deduce that the hardening is mainly governed by stellar interactions that originate from the NSCs and the bulge contribution would not be significant.
	
	\subsection{NGC205} \label{subsec:results205}
	
\begin{figure}
		\centerline{
			\resizebox{0.95\hsize}{!}{\includegraphics[angle=270]{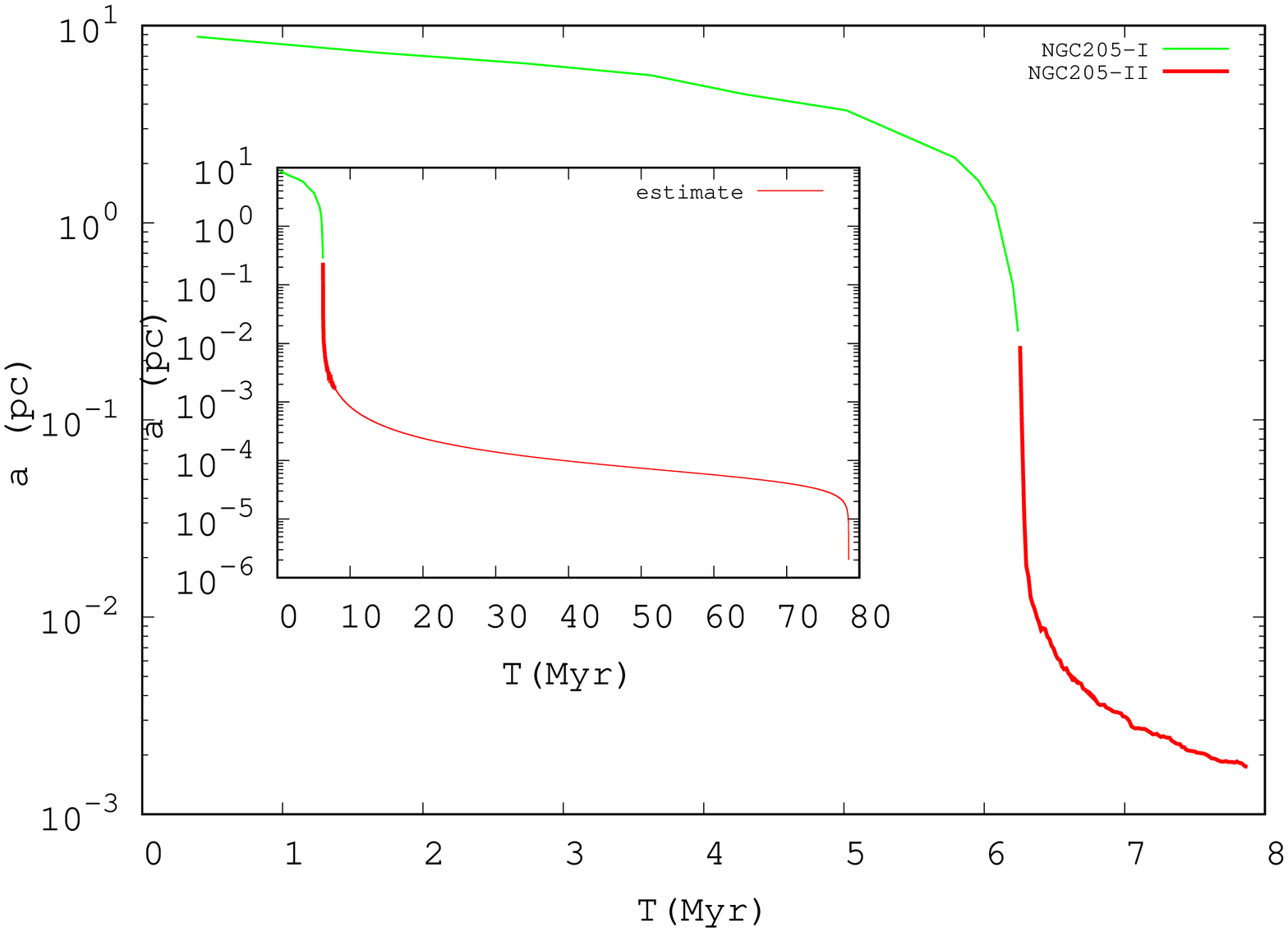}}
		}
		\centerline{
			\resizebox{0.95\hsize}{!}{\includegraphics[angle=270]{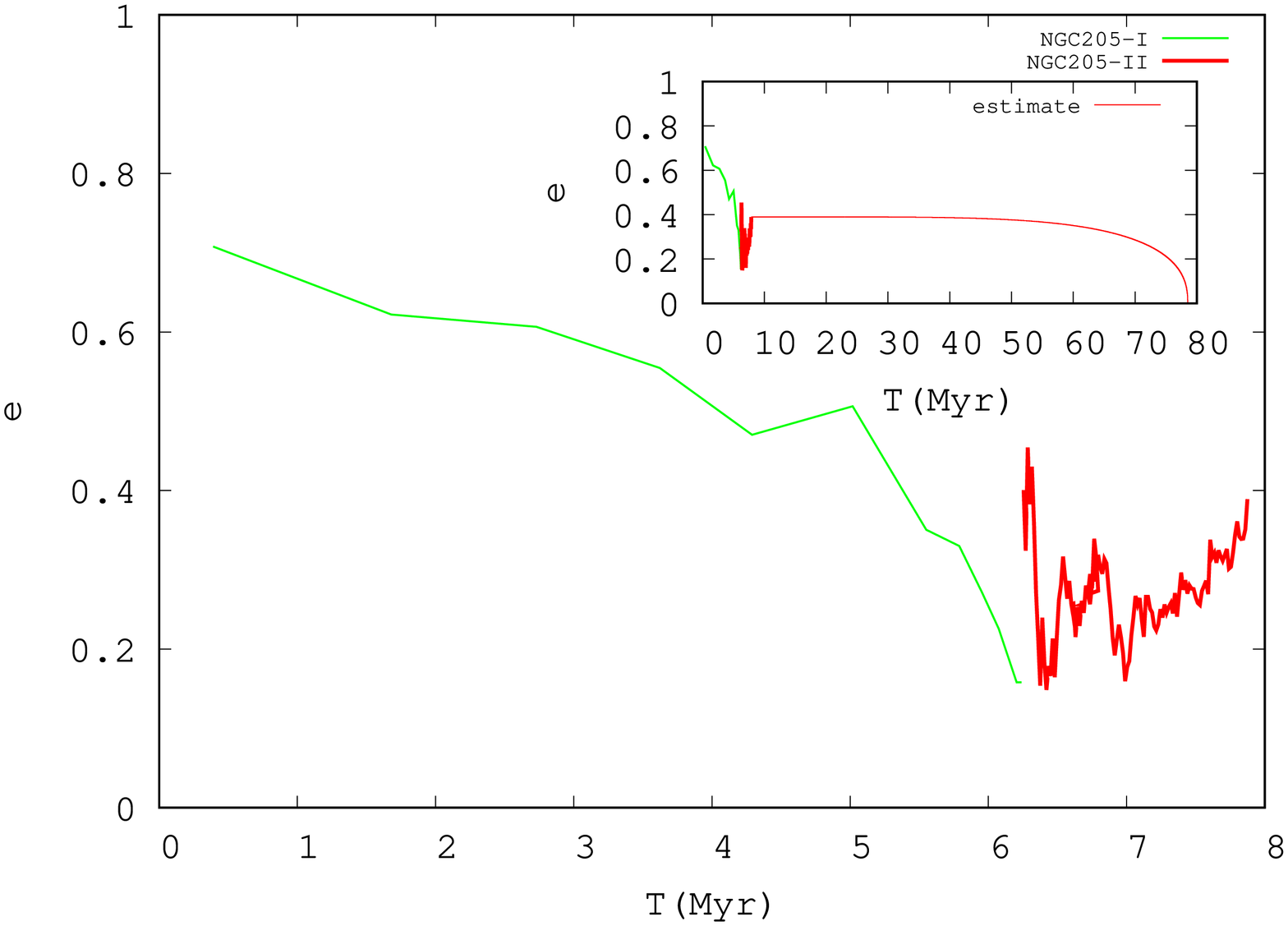}}
		}
		\caption{
			IMBH pair evolution for NGC 205. The inset panels show the estimated evolution for a constant eccentricity after the simulation stops. 
		} \label{fig:ngc205param}
	\end{figure}

	The mass of the central black hole in our model of the NGC 205 nucleus is $4 \times 10^4 M_{\odot}$, as discussed in section \ref{subsec:ngc205}, which pushes the numerical limits of our simulation. We introduce an equal mass secondary at an initial distance of 15 pc with $50\%$ of circular velocity. For this particular run, we doubled the particle number and reduced the stellar-black hole interaction softening to $7 \times 10^{-5}$ pc, which allowed us to better resolve the smaller black hole mass and influence radius. The IMBH binary evolution is presented in figure \ref{fig:ngc205param}. A Keplerian binary forms around $T_b \sim 6$ Myr 
	with low eccentricity $\sim 0.2$ and grows gradually, but we could not follow the binary evolution as long as in the other runs, due to the extremely high computational cost for integrating a hard binary with 4 million particles with a milliparsec separation. We estimate the binary IMBH hardening rate by fitting a straight line to $1/a$ during the last million years of binary evolution and find $s \sim 299 {\rm pc}^{-1}$/Myr. This is the highest hardening rate out of all galaxies in our sample thanks to the highest stellar density at the influence radius and the smallest IMBH binary mass. Subsequent evolution to coalescence is estimated using the formalism described in section \ref{subsec:reciepe} using a constant value of eccentricity ($e = 0.39$) at the end of our run. 
	The IMBHs merge rather swiftly in 79 Myr for a constant $e$ scenario. 
	
\begin{figure}
		\centerline{
			\resizebox{0.95\hsize}{!}{\includegraphics[angle=270]{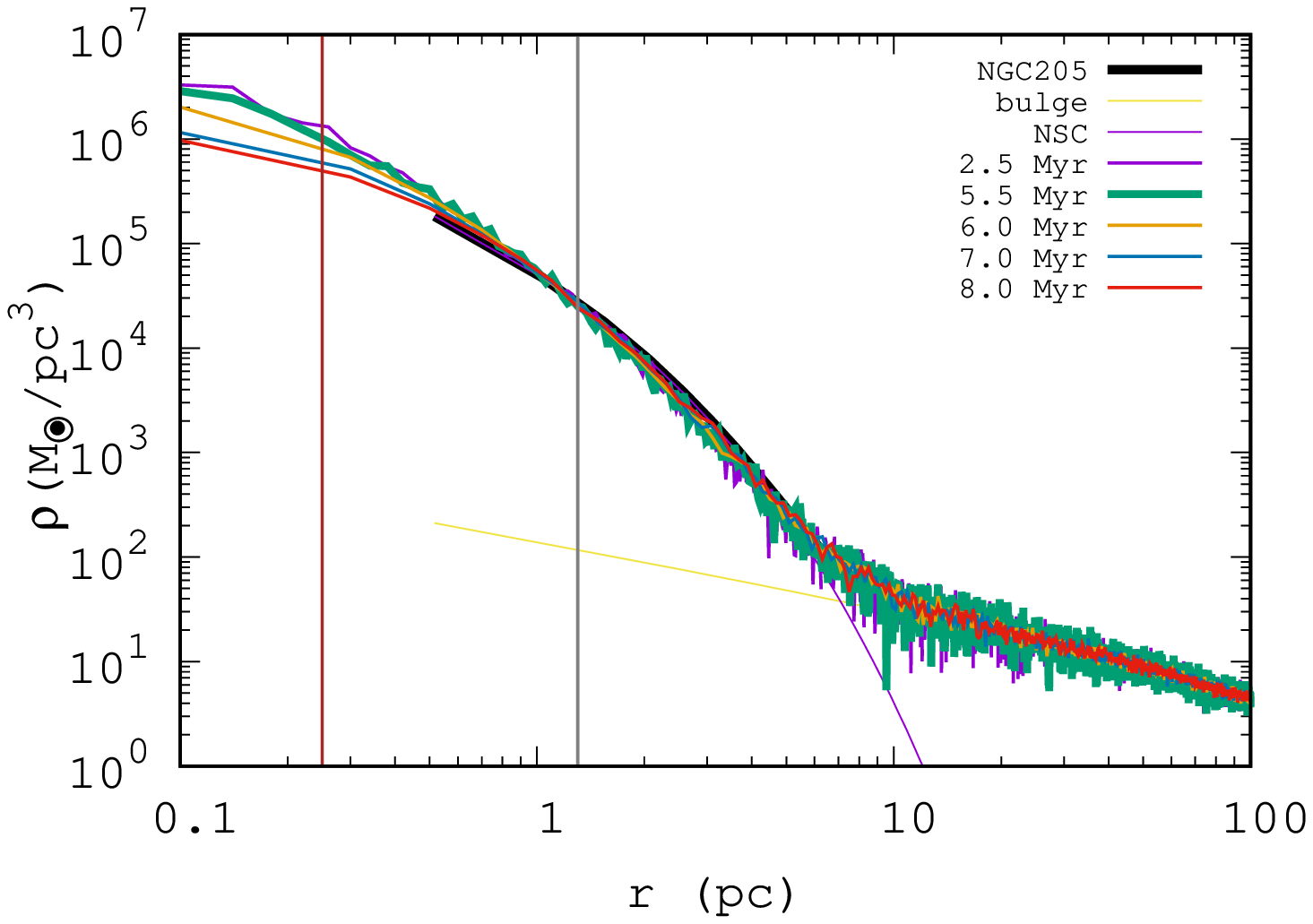}}
		}
		\centerline{
			\resizebox{0.95\hsize}{!}{\includegraphics[angle=270]{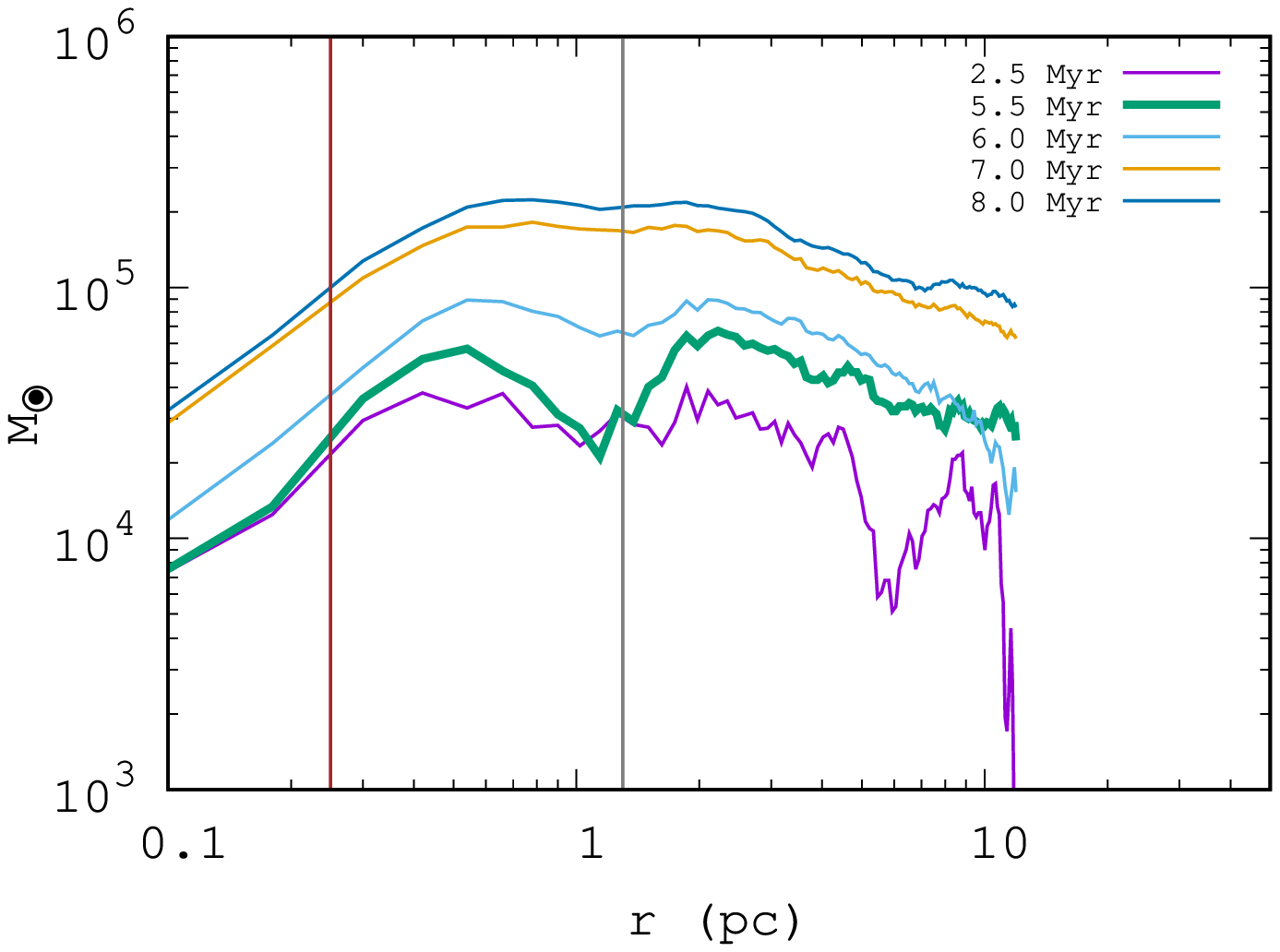}}
		}
		\caption{
			Density profiles (top) and cumulative mass deficits (bottom) for the NGC205 run at various times (in Myr). Thick green curve marks binary formation time. As earlier, the vertical brown line is the IMBH binary's influence radius, and the vertical grey line marks the effective radii of the NSC.
		} \label{fig:ngc205dm}
	\end{figure}

	Core scouring by the IMBH binary has not depleted the center of galaxy as much as in other runs; this is mostly due to the smaller mass binary, though the incomplete run likely played a role as well. The damage to the central density profile does not extend far beyond the effective radius of the NSC (figure \ref{fig:ngc205dm}). The influence radius of IMBH binary extends to 0.25 parsec, which is slightly greater than initial influence radius of the central IMBH. We also notice that NSC stellar density at the influence radius is still more than two orders of magnitude greater than that of bulge. This finding continues to suggest that for such low mass binaries, the majority of stellar encounters originate from the NSC even in the hard binary phase. Total mass deficit induced by IMBH binary at the center of our run is $4.4 \times M_{IMBHs}$.
	
	\begin{table*}
		\begin{center}
			\vspace{-0.5pt}
			\caption{IMBH Binary and Host Parameters and Merger Results} 
			\begin{tabular}{l c c c c c c c c c}
				\hline
				Run & $M_{\rm BBH}$($10^6$ M$_{\odot}$) & $\rho_{0}$ (M$_{\odot}/{\rm pc^3}$)  & $r_{h}$(pc)  & $\rho_{r_h}(M_{\odot}/{\rm pc^3})$  & $v/\sigma$ & $M_{\rm NSC}/M_{\rm BBH}$ & $s( {\rm pc^{-1}/Myr})$ & $e_{\rm f}$ & $T_{\rm coal} ({\rm Myr})$\\
				\hline
				M32-ret & $3.0$ & $10^7$ & $6.5$ & $10^4$ & $0.7$ & $4.83$ & $1.94$ & $0.91$ & $182$ \\
				M32-pro & -- & -- & -- & -- & -- & -- & $1.77$ & $0.09$ & $506$ \\
				NGC5102-ret & $1.76$ & $10^6$ & $1.9$ & $4.46 \times 10^4$  & $0.6$ & $37.0$ & $8.3$ & $0.99$ & $23$ \\
				NGC5102-pro & -- & -- & -- & -- & -- & -- & $8.8$ & $0.54$ & $178$ \\
				NGC5206 & $0.94$ & $10^6$ & $2.9$ & $2 \times 10^4$ & $0.25$ & $15.4$ & $3.73$ & $0.4$ & ($531,178$) \\
				NGC404 & $0.14$ & $10^5$ & $0.7$ & $1.1 \times 10^5$ & $0.2$ & $103.0$ & $49$ & $0.5$ & $251$ \\
				NGC205 & $0.08$ & $10^6$ & $0.25$ & $7 \times 10^5$ & $0.2$ & $33.75$ & $299$ & $0.39$ & $79$ \\
				\hline
			\end{tabular}\label{tab:resultsim1}
			\vspace{15pt}
			
			Column~1: Galaxy run. Column~2: Black hole binary mass. Column~3: Central stellar density. Column~4: IMBH binary influence radius. Column~5: Stellar density at the influence radius. Column~6: rotational velocity to velocity dispersion ratio. Column~7: Nuclear star cluster to IMBH binary mass ratio. Column~8: IMBH binary hardening rate. Column~9: IMBH binary eccentricity at the end of our simulations. Column~10: Estimated merger time.
		\end{center}
		\vspace{15pt}
	\end{table*}
	
	\section{Summary and Conclusion}
	
	A large motivation for modeling the dynamics of binary black holes within galaxy centers is the prospect of gravitational wave detection from space by LISA \citep{amaro+17}. However, due to numerical issues, many prior studies, including some of ours, are forced to model scenarios that are quite different than what is expected for LISA sources. For example, to resolve the dynamics between black holes and the stellar loss cone, it is typical to deploy unphysically massive black holes compared to the stellar mass, and to speed the calculation, it is common to both flatten and lower the stellar density. Many prior simulations also neglect to include galaxy components, and/or launch the secondary black hole close to the galaxy center, and/or assume equal mass mergers even though this is thought to be rare. This study pushes the numerical limits to attempt a higher degree of realism for LISA sources. We focus on evolution of IMBH pairs in the mass range $10^4-10^6~ M_{\odot}$; mergers of these lower mass black holes lie near the peak of the LISA sensitivity window. Importantly, black holes in this mass range dwell in galaxies with key differences from massive hosts: they are dwarf galaxies with rotating and incredibly dense nuclear star clusters. As dwarf galaxies dominate in number, even a relatively small occupation fraction of such binaries may contribute significantly to the LISA event rate. 
	
	While there has been much attention paid to the evolution of binary SMBHs in early and late type galaxy hosts, there is not yet a single study following the dynamics of IMBHs in the centers of dwarf galaxies. We show that the dynamics and evolution of IMBH pairs depends critically on structure and kinematics of their stellar surroundings. Of importance are the stellar density and velocity dispersion at the influence radius \citep{seskha+15}. However, rotation and flattening, if present, can affect the binary shrinking rate, the eccentricity both at formation and late binary evolution times, and the center of mass motion of the binary itself \citep{kha20}. These quantities are relatively easier to obtain from observations of massive galaxies as they have comparatively larger and more luminous nuclear regions than do dwarf galaxies. However, recent deep and integral-field observations have begun to map the central structure and kinematics of dwarf galaxies  \citep{seth08,seth10,den14,ngu18}, and have revealed a range of features capable of impacting IMBH pair evolution in these systems.

	\begin{figure}
		\centerline{
			\resizebox{0.95\hsize}{!}{\includegraphics[angle=270]{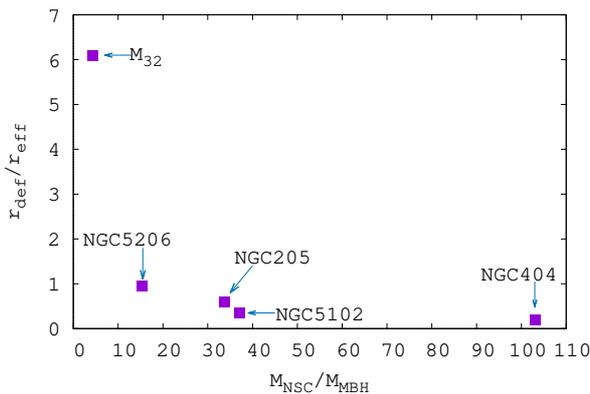}}
		}
		\caption{
			Radius at which the mass deficit peaks, in units of the NSC effective radius, versus the mass of the NSC, in units of IMBH mass. 
		} \label{fig:nscmbh}
	\end{figure}
	
	We build galaxy models of nearby dwarfs with IMBH candidates whose detailed observations provide the key parameters to construct kinematic and structural analog models \citep{ngu17,ngu18}. We study the evolution of similar mass IMBH pairs in these systems using high resolution direct $N$-body simulations. Some of the key findings of our studies are summarised as below.
	
	IMBH pair separation shrinks very rapidly from 50 pc down to a parsec in about 5-20 Myr due to efficient dynamical friction in the high stellar density background provided by NSCs. This shows that if IMBHs can be brought inside central 100 pc either by globular clusters or by dwarf galaxy mergers they will quickly form a binary.
	
	Once in a binary, the IMBH binary hardening rates scale well with the stellar density at the influence radius and inversely with IMBH binary mass, as suggested by table \ref{tab:resultsim1}, although rotation also plays a role by effecting the eccentricity. The highest hardening rates occur in NGC205, which has the largest density at the binary influence radius, as well as the smallest IMBH binary mass of all in the suite.  In contrast, the lowest hardening rate occurs in the M32 nuclei; although it as the highest central density, the cuspy density profile results in a comparatively low stellar density at the IMBH binary influence radius. Additionally, the IMBH mass in the M32 nucleus is the highest in our sample. 
	
	Eccentricities in the hard binary regime have intermediate to very high values $0.4-0.99$. The largest eccentricities occur for retrograde orbits in NGC5102 and M32. NGC5102 is only mildly rotating, while M32 possesses a strongly rotating nucleus. In case of M32, the binary flips its orientation soon after it becomes a binary. This causes the eccentricity to reduce slightly from 0.98 to 0.91 after the binary plane flips and the IMBH binary evolves in co-rotation with its stellar surroundings. For NGC5102, mild rotation is not enough to flip the IMBH plane and it enters the hard binary regime still counter-rotating with its stellar surroundings; the resulting eccentricity growth approaches unity. This is consistent with our earlier findings reported in \citet{kha20}. We estimate residual eccentricities that may persist into the LISA regime; however we plan to follow up with more detailed simulations evolving well into the gravitational radiation regime to confirm this estimate, as it has important implications for LISA waveforms and detection.
	
	The gravitational slingshot effect during the 3-body scattering phase results in the partial disruption of the NSC, expanding the sphere of influence to larger radii. Our analysis suggests that for most of the galaxy models in our suite (with the exception of M32), the majority of stellar encounters originate from the NSC.  Table \ref{tab:resultsim1} shows the NSC-to-IMBH binary mass ratio for our suite. Note that this ratio is greater than 10 for all systems other than M32. In figure \ref{fig:nscmbh}, we explore the radius to which the IMBH binary depletes the galaxy as a function of this NSC-to-IMBH mass ratio.  This figure suggests that for systems having a high NSC-to-binary mass ratio, the IMBH binary evolution is predominantly driven by NSCs, as the mass deficit is confined within the effective radius of the NSC. Conversely, a relatively more massive IMBH binary would require a stellar reservoir larger than the NSC. 
	
	For commensurate mass IMBH binaries, merger times are of the order of a few hundred Myrs, which may imply high IMBH merger rates for LISA. Factors defining this evolution are stellar density at the influence radius, velocity dispersion, eccentricity, and rotation. Our results are in contrast to \citep{ogi19} who found merger times of order a Gyr or longer for similar mass IMBHs in their study of merging NSCs, and further found that the merger time strongly depended on the IMBH mass ratio. However, the NSC-to-IMBH binary mass ratio in their merger suite is 5 for equal mass binaries, which from our work implies that the NSC does not contain a significant enough stellar reservoir to merge the IMBHs. In their work, they neglect including a bulge and hence the equal mass binaries take a long time to merge. As the mass ratio increases in their work, the NSC-to-IMBH binary ratio approaches 10, and their merger times begin to be consistent with ours. We argue that this is due less to a new dynamical process that depends strongly on mass ratio and more on the fact that the 3-body scattering physics is better resolved.
	
	Since most low mass galaxies contain NSCs, and since we find that IMBH binary mergers disrupt them, it begs the question as to whether any nucleated dwarf has experienced a recent IMBH merger. Indeed, one channel of NSC-formation involves purely stellar assembly from the inspiral and disruption of globular clusters, and in this case, cusp-rebuilding timescales are expected to be long~\citep{Antonini13,Gnedin14}. However, the presence of young star formation in many NSCs (particularly within late-type galaxies) point to a second channel of NSC formation: in-situ star formation, which argues for a copious inflow of gas and a rebuilding timescale of $10^8$ years~\citep{Seth06,Walcher06}.  In addition, highly unequal mass IMBH mergers would presumably have a less disruptive effect on the NSC; simulating such as system would require a significant increase in particle number and remains for future work. Taken together, we speculate that the mere presence of a NSC does not rule out the possibility that the galaxy could also be a potential host of a LISA source.  
	

	

	\section*{Acknowledgments}
	
	We thank Alister Graham for useful discussions. We acknowledge the support by Vanderbilt University for providing access to its Advanced Computing Center for Research and Education (ACCRE). 
	FK and KHB were supported through NASA ATP Grant 80NSSC18K0523.
	
	\section*{Data Availability Statement}
	
	The data underlying this article will be shared on reasonable request to the corresponding author.
	
	

	\bibliographystyle{mnras}
	\bibliography{ms}

	\bsp	
	\label{lastpage}
\end{document}